\documentclass[openacc]{rstransa}

\newtheorem{theorem}{\bf Theorem}[section]

\usepackage{enumitem}
\usepackage{mathrsfs}
\usepackage{bm}
\usepackage{amsmath}
\usepackage{amsthm}
\usepackage{amssymb}
\usepackage{stmaryrd}
\usepackage{mathtools}

\usepackage[capitalize]{cleveref}

\usepackage{color}
\RequirePackage[dvipsnames,usenames]{xcolor}

\newcommand{\eq}[1]{\begin{align}#1\end{align}}

\newcommand{\II}{{\mathcal{I}}}
\newcommand{\SSSS}{{\mathbb{S}}}
\newcommand{\SSS}{{\widehat{\sum}}}

\newcommand{\AAA}{{\mathcal{A}}}
\newcommand{\BBB}{{\mathcal{B}}}

\newcommand{\E}{{\mathbb{E}}}

\newcommand{\R}{{\mathbb{R}}}
\newcommand{\B}{{\mathbb{B}}}
\newcommand{\N}{{\mathbb{N}}}

\newcommand{\NN}{\mathcal{N}}
\newcommand{\MM}{\mathcal{M}}
\newcommand{\LL}{\mathcal{L}}
\newcommand{\PP}{\mathcal{P}}

\newcommand{\CC}{\mathcal{C}}

\newcommand{\pa}{{\rm{pa}}}

\newcommand{\family}{{\rm{fa}}}
\newcommand{\desc}{{\rm{desc}}}

\newcommand{\dSt}{{\langle\dot{\sigma}^{\NN}(t)\rangle}}

\newcommand{\dSot}{{\langle\dot{\sigma}^\omega(t)\rangle}}

\newcommand{\oo}{{\omega}}

\newtheorem{example}{Example}

\newtheorem{proposition}[theorem]{Proposition}

\begin{document}

\title{Strengthened second law for multi-dimensional systems coupled to multiple thermodynamic reservoirs}

\author{
David H. Wolpert$^{1}$
}

\address{$^{1}$Santa Fe Institute, New Mexico, USA\\
Complexity Science Hub, Vienna\\
Arizona State University, Tempe, Arizona\\
International Center for Theoretical Physics, Italy\\
\texttt{http://davidwolpert.weebly.com}
}

\subject{Emergence, Non-equilibrium statistical physics}


\corres{David H. Wolpert\\
\email{david.h.wolpert@gmail.com}}

\begin{abstract}
The second law of thermodynamics can be formulated as a restriction on the evolution of the 
entropy of any system undergoing Markovian dynamics.
Here I show that this form of the second law is strengthened for multi-dimensional, complex systems,
coupled to multiple thermodynamic reservoirs, if we have a set of \textit{a priori} constraints
restricting how the dynamics of each coordinate can depend on the other coordinates.
As an example, this  strengthened second law (SSL) applies to complex systems composed of
multiple physically separated, co-evolving subsystems, each identified as a coordinate of the overall system.
In this example, the constraints concern how 
the dynamics of some subsystems are allowed to depend on the states of the other subsystems.
Importantly the SSL applies to such complex systems even if some of its subsystems can change state simultaneously,
which is prohibited in a multipartite process. 
The SSL also strengthens previously derived
bounds on how much work can be extracted from a system using feedback control, if the system is multi-dimensional.
Importantly, the SSL does not require local detailed balance.
So it potentially applies to complex systems ranging from interacting economic agents to co-evolving biological species.
\end{abstract}

\maketitle

\section{Introduction}

Statistical physics concerns experimental
scenarios where we have restricted information concerning the state of a system $x \in X$, which is quantified as a probability distribution 
over those states, $p_x(t)$. In particular, the recently developed variant of statistical physics 
called ``stochastic thermodynamics'' concentrates on systems that evolve
according to a continuous time Markov chain (CTMC). For a countable state space, this means that $p_x(t)$ evolves according
to a linear differential equation,
\eq{
\dfrac{d p_x(t)}{dt} = \sum_{x'} K^{x'}_x(t) p_{x'}(t)
\label{eq:0}
}
(Note that the rate matrix $K(t)$ can depend on time $t$.)



\maketitle

Analyzing systems which evolve according to~\cref{eq:0} has led to formulations of the 
second law of thermodynamics which apply even if the system 
is evolving while arbitrarily far out of thermal equilibrium~\cite{van2015ensemble,seifert2012stochastic}.
If we apply one of these formulations of the second law to any system evolving according to \cref{eq:0} 
while coupled to a single (infinite) heat
bath at temperature $T$, and assume that the rate matrix is related to
an underlying Hamiltonian via \textbf{local detailed balance} (LDB), we get
\eq{
\dfrac{Q}{T} \le \Delta S
\label{eq:1}
}
where $Q$ is the total heat flow into the system from its heat bath during the dynamics,
and $\Delta S$ is the change in Shannon entropy of the system during the process. 

If LDB does not hold, \cref{eq:1} will not hold either, if we wish to interpret $Q$ as 
thermodynamic heat flow. However, for any rate matrix,
regardless of whether it obeys LDB, 
\eq{
\int_{t_i}^{t_f} dt  \sum_{x'} K^{x'}_x(t) p_{x'}(t) \ln \dfrac{K^{x}_{x'}(t)}{K^{x'}_{x}(t)} \,  \le \Delta S
\label{eq:2}
}
(for a process lasting from time $t_i$ to $t_f$). The quantity on the LHS of \cref{eq:2} is
called the total expected \textbf{entropy flow} (EF) into the system during the process. The difference
between the entropy change of the system (the RHS of \cref{eq:2}) and the EF is called the \textbf{entropy production} (EP),
written as $\sigma$. So \cref{eq:2} can be re-expressed as
\eq{
\sigma \ge 0
\label{eq:2a}
}
Crucially, the inequality \cref{eq:2a} holds for \textit{any} CTMC, even 
a CTMC that has no thermodynamic interpretation,
i.e., a CTMC which models a process that does not involve energy transduction. 
So \cref{eq:2a} applies to dynamic models of everything from stock markets to the evolution of the joint state
of an opinion network, so long as those models are CTMCs.

In many experimental scenarios, while we are restricted in the information we have concerning 
the system's state, we also have some other information that does not directly concern the
system's state, in the form of conditions satisfied by the dynamics 
of the system. Recently \cref{eq:2a} has been strengthened, by adding non-positive terms to the RHS
that incorporate this kind of information concerning the dynamics. Examples of these new results include
``thermodynamic uncertainty relations'' 
(TURs~\cite{horowitz_gingrich_nature_TURs_2019,liu2020thermodynamic,koyuk2020thermodynamic,falasco2020unifying}), ``speed limit theorems'' (SLTs~\cite{shiraishi_speed_2018,zhang2018comment,okuyama2018quantum,van2020unified,gupta2020thermodynamic}),
``thermodynamic first passage bounds''~\cite{neri2017statistics,gingrich2017fundamental,falasco2020dissipation,roldan2015decision,neri2021dissipation}, etc. 

Unlike \cref{eq:2a} though, these bounds require measuring variables as they change during the process,
in addition to knowing the beginning and ending distributions, $p_x(t_i)$ and $p_x(t_f)$. (For example, TURs rely on
measuring accumulated currents, and speed limit theorems rely on measuring integrated activity.) This 
limits their experimental applicability.

In this paper, I derive new strengthened forms of \cref{eq:2a} that, like the TURs and speed limit
theorems, incorporate information concerning the dynamics of the system. However, unlike the TURs,
speed limit theorems, etc., these new strengthened forms of \cref{eq:2a} do \textit{not} require
measuring variables as they change during the process.

These strengthened forms of \cref{eq:2a} apply whenever we have information about
which of the coordinates of the system can have their dynamics directly depend on which of the other 
coordinates. Formally, such information takes the form of constraints on the 
rate matrix $K(t)$ of the CTMC governing the dynamics of the system. (See also~\cite{kolchinsky_wolpert_2020_thermo_under_constraints}.)
I call this kind of restriction on the allowed dynamics a ``dependency constraint''.

\begin{figure}[tbp]
\includegraphics[width=75mm]{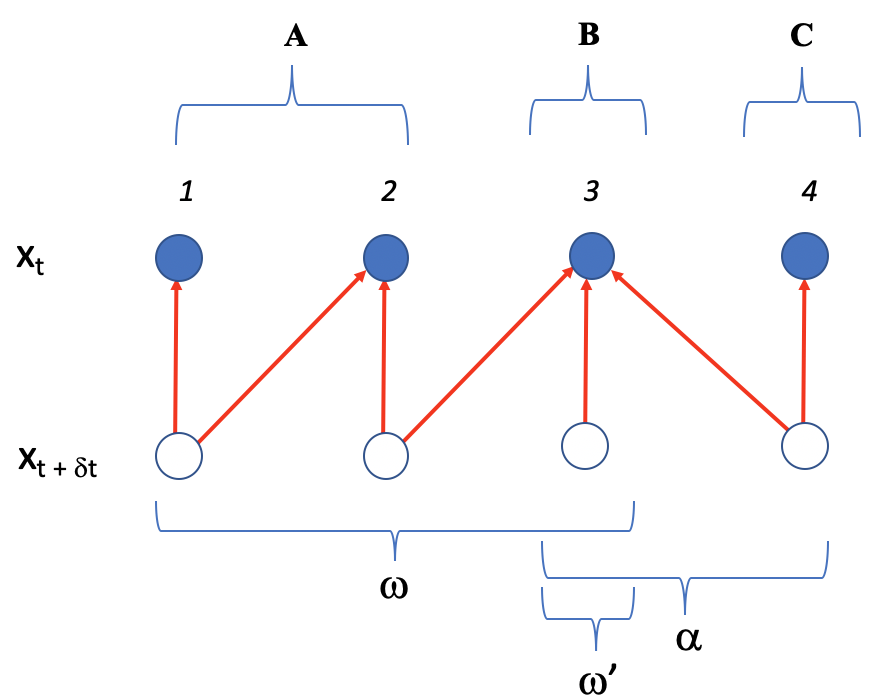}
\caption{Four interacting subsystems, $\{1, 2, 3, 4\}$, grouped into three sets, $\{A, B, C\}$.
The red arrows indicate dependencies in rate matrix of the overall system. So for example
$B$ evolves autonomously, but is continually observed by $A$ and $C$. 
(The implicit assumption that $B$ is not affected by the back-action of the observation
holds for many real systems such as colloidal particles and macromolecules~\cite{sagawa_nonequilibrium_2012}.) 
Note that the statistical coupling
between $A$ and $C$ could grow with time, even though the rate matrix does not directly
couple their dynamics. The three overlapping sets indicated at the bottom of the figure specify the three units of a unit structure
for this process, as discussed in the text. As an illustration of some of the definitions below, there is
one reservoir coupled to the system that has subsystem $2$ as its puppet set, with both subsystems $2, 3$ as its leader set.
}
\label{fig:1}
\end{figure}

As an example, 
consider a random walker over a two-dimensional finite lattice, $Y_1 \times Y_2$.
For simplicity take $|Y_1| = |Y_2| = L N$ for two positive integers $L, N$. The lattice is 
coarse-grained into a set of $N^2$  non-overlapping squares each of size $L \times L$, 
and the position of the walker in the lattice is represented three-dimensionally, 
by a pair of coordinates $X_1 = \{1, \ldots, L\}, X_2 = \{1, \ldots, L\}$ and an integer $X_3 \in \{1, \ldots, N^2\}$. (The value $x_3 \in X_3$ specifies the precise coarse-grained square, 
while $(x_1, x_2) \in X_1 \times X_2$ specifies the coordinates within that square.)
In addition to position in the lattice, the walker has internal stores of two nutrients, $A$ and $B$,
specified (up to some coarse-graining) by values in the finite sets $X_A$ and and $X_B$, respectively. 
So the state space of the walker
is $X = X_A \times X_B \times \prod_{i=1}^3 X_i $, 
i.e., 
those five variables are the five coordinates of the walker.

We can suppose that both $x_1$ and $x_2$ evolve autonomously, 
independently of all other variables, according to two associated rate matrices, i.e.,
the walker engages in two independent random walks, one in each of the two
directions across the lattice. Note though that $x_3$'s dynamics
will depend on $x_1$ and $x_2$ in general, and that there will sometimes be simultaneous transitions of $x_3$ and
some other coordinate.
For example, suppose $x_1 = L$, so the walker is at the extreme value of $X_1$ within some square, adjacent
to the next coarse-grained square. Suppose as well that in the next step, 
the walker moves into that adjacent square. So simultaneously $x_1$ changes to $1$ while $x_3$ must 
also change, since the coarse-grained square changes. However,
for other changes in $x_1$, $x_3$ remains unchanged.

We can also suppose that the dynamics of $x_A \in X_A$ depends only on the walker's current position in $X_1$ and their
current amount of $x_A$, i.e., it depends only on $(x_A, x_1)$. Similarly, the
dynamics of $x_B \in X_B$ depends only on $(x_B, x_2)$. (For example, this would be the case if densities of those two
nutrients were arranged appropriately across the lattice, and the walker at a given location accumulates those nutrients based
on their densities at that location.) Summarizing, the dependency constraints are that $x_A$ depends only on 
$x_1$ (in addition to depending on its own state), $x_B$ depends only on $x_2$ (in addition to its own state), 
$x_1$ and $x_2$ are autonomous, while $x_3$ can depend on $x_1$ and  / or $x_2$ (in addition to itself).

There are many other dynamic processes over a state space which obey a set of dependency constraints among the
coordinates of the state space, but those
coordinates do not specify characteristics of a single agent like a walker traversing a lattice.
For this special issue on  the topic of `Emergence',
perhaps the most important type of system that evolves subject to dependency constraints
is a system that comprises a set of physically separated subsystems, co-evolving with one another,
%
with each subsystem's state being identified as a different coordinate~\cite{horowitz_multipartite_2015,wolpert2020minimal,wolpert2020fluctuation}.
%
In this kind of system, 
dependency constraints governing the dynamics of each coordinate, specifying which other 
coordinates can directly affect its dynamics,
amount to constraints on the dynamics of each subsystem, specifying which other subsystems can directly affect
its dynamics.
As a concrete illustration, consider the scenario investigated in~\cite{hartich_sensory_2016,bo2015thermodynamic},
in which receptors in the wall of a cell sense the concentration of a ligand in the intercellular medium,
and those receptors are in turn observed by a ``memory'' subsystem inside the cell. Modify this scenario by introducing a second
cell, which is observing the same external
medium as the first cell. Assume that the cells are far enough apart physically so that
their dynamics are independent of one another. This gives us the precise scenario in \cref{fig:1}, where 
subsystem $3$ is concentration in the external medium, subsystem $2$ is the state of the receptors of 
the first cell, subsystem $1$ is the memory subsystem of the first cell,
and subsystem $4$ is the state of the receptors of the second cell. 

In this paper I consider how a set of dependency constraints on an evolving system affect its thermodynamics. 
My main result shows how such a set of dependency constraints can strengthen \cref{eq:2a}, by adding
an expression to its RHS. This expression involves only those dependency constraints and the starting and ending distribution of the system.
%
As a caveat, this new lower bound on EP is not always positive, i.e., it is not always stronger than the conventional second law, \cref{eq:2a}.
However, I show below that for any
set of dependency constraints, there is a conditional distribution $p\left(x(t_f) \, | \, x(t_i)\right)$ that can be implemented
by a rate matrix obeying those constraints, together with an initial
distribution $p(x(t_i))$,
such that \textit{every} rate matrix that implements that conditional distribution must
result in a non-negative EP when applied to that initial distribution.
Indeed, for some sets of dependency constraints, this new
EP bound is stronger than the conventional second law {no matter what} $p(x(t_i))$ 
and $p\left(x(t_f) \, | \, x(t_i)\right)$ are (so long as $p\left(x(t_f) \, | \, x(t_i)\right)$ is consistent with
the dependency constraints).

Some of the TURs,  SLTs, etc., rely on the dynamics obeying LDB. LDB is not required for the
new extension of the second law derived here. This means that (for example) this new extension applies
to multipartite systems that have ``directed'' (sometimes called ``non-reciprocal'') interactions rather than undirected interactions 
among the subsystems, i.e., interactions in which there is exactly zero 
back-action~\cite{hartich_stochastic_2014,hartich_sensory_2016,sagawa2009minimal,parrondo2015thermodynamics,verley_work_2014,lipowski2015phase,sanchez2002nonequilibrium,loos2020irreversibility}.
Very often, these systems violate strict LDB, and so their thermodynamic analyses are, at best,
approximations. (See discussion in appendix in~\cite{wolpert2020fluctuation} of some conditions that
justify this approximation.) In contrast, the result derived below
applies exactly to any scenario where there is no back-action, with no approximation. 
In addition, this result holds even if the dynamics allows multiple
coordinates to change simultaneously. In particular, in the special case that each coordinate is a separate subsystem,
the result does not require that the dynamics be a multipartite process~\cite{horowitz_multipartite_2015}. 

Due to these relaxations of the assumptions made in conventional stochastic thermodynamics, 
the results below are not restricted to thermodynamic systems, involving energy
transduction. The results hold for any CTMC, even if the rate matrix 
does not reflect physically coupling between the system and one or more external thermodynamic 
reservoirs, as it does in conventional applications of stochastic thermodynamics~\cite{van2015ensemble}. 

However, the strengthened second law derived below has special physical significance
in the common scenario where the dependency constraints arise because the system's dynamics is
governed by coupling with external reservoirs, and there are restrictions on that coupling.
For example, a common physical scenario is where the
system has multiple subsystems, and each subsystem is coupled to a physically distinct part of
a shared reservoir. Due to the physical separation of those parts of the reservoir, each connected to a different
subsystem, the usual assumption of time-scale separation between the dynamics of the overall system and that of the
reservoirs means that the different subsystems are effectively coupled to independent reservoirs from one another.\footnote{It is 
important to note that this separation is between the time-scale of the dynamics of the overall system and the time-scale of the implicit 
dynamics of the thermodynamic reservoirs that are coupled to the system~\cite{seifert2012stochastic}. It does \textit{not} concern
the time-scales of the dynamics of the different coordinates of the system. For analysis of the latter kind of time-scale
separation in the special case of a bipartite system, see~\cite{busiello2020coarse}, and 
for a more general analysis, see~\cite{strasberg2017stochastic}.}
Such systems evolve as a multipartite process (MPP), in which no transitions
are allowed in which more than two subsystems change their states exactly simultaneously~\cite{horowitz_multipartite_2015}. If the
system is an MPP, and the dynamics of each subsystem obeys LDB, then we can use stochastic thermodynamics
to identify various attributes of that dynamics with experimentally measurable thermodynamic quantities~\cite{horowitz2014thermodynamics,horowitz_multipartite_2015,hartich_stochastic_2014,barato2014stochastic,Brittain_2017}.
More generally, there are systems with multiple coordinates that aren't usually viewed as separate ``subsystems'', but
where the global dynamics arises due to the system's coupling with thermodynamic reservoirs, and where each reservoir is only 
coupled to a single coordinate. These systems can also be modeled as MPPs, and analyzed accordingly.

Generalizing further, there are other kinds of systems that also
have multiple coordinates, where the global dynamics arises due to the system's coupling 
with thermodynamic reservoirs, just like in an MPP. Also like in an MPP, 
each reservoir in these systems  is only coupled to a proper subset of the coordinates, which
results in dependency constraints. In contrast to an MPP however, some reservoirs are coupled to more than one coordinate.
As an example, as stated in~\cite{chetrite2019information}: ``Fluctuations in biochemical networks, e.g. in a living cell, have
a complex origin that precludes a description of such systems in terms of
bipartite or multipartite processes, as is usually done in the framework of
stochastic and/or information thermodynamics.'' The strengthened second law I present below applies to these
generalized forms of MPPs as well as to MPPs.

In the next section I formalize dependency constraints as restrictions on the rate matrix of  a CTMC.
This is followed by a section in which I 
use this formalization to derive an expression for the EP of system that involves the triple of
\{the rate matrix dependency constraints,
the initial distribution over states, the final distribution over states\}, together with certain other factors. 
In the following section I derive a lower bound on that expression for EP which depends only on
the triple of \{dependency constraints, initial distribution,
final distribution\}, without those other factors. In particular, this lower bound does not depend on any
properties of the rate matrix, other than the dependency constraints. This lower bound is my main result.
In the following section this main result to analyze how the thermodynamics of feedback control~\cite{sagawa2009minimal,parrondo2015thermodynamics,kolchinsky_wolpert_2020_thermo_under_constraints} 
changes when we 
know that the system being controlled obeys a given set of dependency constraints. In the following section I present
a set of examples of my main result. I end with some discussion, in particular of the relation of the new result to 
other results in the literature.


\section{Rate matrix unit structures}

I begin by defining notation. First, I write the state space of the system as $X = \prod_{i=1}^N X_i$,
where each finite state space $X_i$ is a coordinate of the system. I write the set of $N$ coordinates as $\NN$.
As examples, each coordinate could specify the state of a physically separate subsystem of the overall system, 
or it could specify a position on one axis of a lattice, or it could indicate a degree of freedom in a multi-scale
specification of the state of the system. 

The distribution $p_x(t)$ over the states of the system is assumed to evolve according
to a continuous time Markov chain (CTMC)\footnote{Note that assuming the state space of the system is a 
Cartesian product does not limit the applicability of the analysis. Suppose that
the set of physically allowed states are a subset of such a Cartesian product, $Y \subset \prod_i  X_i$, but that $Y$
is not itself such a Cartesian product. We can model such a scenario using the Cartesian product state space $X = \prod_i X_i$, simply
by restricting the rate matrix of the CTMC so that there is zero probability of going from a state in $Y$ to a state in $X \setminus Y$.}.
For any $A \subset \NN$, I write $-A := \NN \setminus A$. So for example, $x_{-A}$ is the vector of all components of $x$ other than those in $A$.
For any set $L$, $\Delta_L$ is the associated unit simplex, and $|L|$ is the number of elements in $L$. In addition, for any function $f(p)$, I write $\Delta f := f(p({t_f})) - f(p({t_i}))$.
The set of bits is $\B = \{0, 1\}$. I write the Kronecker delta as $\delta(a, b)$. For any family of sets, $A = \{a_1, a_2, \ldots\}$,
I define $\cup A = a_1 \cup a_2 \cup \ldots$.

A distribution over a set of values $x$ at time $t$ is written as $p_X(t)$, with its value for
$x \in X$ written as either
$p(x(t))$ or $p_x(t)$, as convenient.
Similarly, I write $p(x(t) \,|\, x({t'}))$ for the conditional
distribution of the state at time $t$ given the state at time $t'$, etc. 
I write Shannon entropy as $S(p_X(t))$, $S_t(X)$, or $S^{X}(t)$, depending on which would result
in the cleanest equations, and write mutual information 
between two random variables $F, G$ as $I(F;G)$.

The distribution over the overall system evolves according to the \textbf{global} rate matrix $K(t)$, as given by \cref{eq:0}.
A \textbf{unit} $\oo \subseteq \NN$ at time $t$ is a set of coordinates such that as the full
system evolves according to $K(t)$, the marginal distribution $p_{x_\oo}$ evolves according to the CTMC
\eq{
\frac{d p_{x_\oo}(t)}{dt} 
	&= \sum_{x'_\oo} K^{x'_\oo}_{x_\oo}(\oo; t) p_{x'_\oo}(t)
\label{eq:15aa}
}
for all $p$, for some associated rate matrix $K(\oo;t)$.
Intuitively, a unit is any set of coordinates whose evolution is independent of the states of the coordinates outside
the unit. Since the dynamics of a unit is given by a self-contained CTMC, all the usual theorems of
stochastic thermodynamics apply to any unit, e.g., the second law~\cite{van2015ensemble}, speed limit theorems~\cite{shiraishi_speed_2018,tasnim2021thermodynamic}, and some of the fluctuation theorems~\cite{jarzynski2000hamiltonian}.

Any union of units is a unit. In addition, it is proven in App.\,A  that any nonempty intersection of units is a unit.
Note that since the dynamics of the full system is a CTMC, \cref{eq:15aa} applies with $\oo$ set
to all coordinates in the system. So $\NN$ is a unit. Note also that in general, the evolution of a coordinate $i$ lying
outside of a unit $\oo$ may depend on the states of coordinates $j$ lying inside $\oo$, even though the reverse is
impossible by definition.

As an example, in \cref{fig:1},
subsystem
$3$ is its own unit, evolving independently of subsystems $2$ and $1$. In contrast, none of the other
three subsystems are their own unit. (For example subsystem $2$'s
dynamics depends on the state of $3$.) 

A set of units defined over a set of coordinates $\NN$ is called a \textbf{unit structure} if it obeys
the following properties~\cite{wolpert2020fluctuation,wolpert2020minimal}:
\begin{enumerate}
\item The union of the units in the unit structure equals all of $\NN$.
\item The unit structure is closed under intersections of its units.
\end{enumerate}
I will generically write any particular unit structure defined over $\NN$ as $\NN^*$.

I will sometimes say that $\NN^*$ \textbf{represents} the set of coordinates $\NN$.
Also, in general for any given rate matrix there are sets of coordinates $\AAA \subset \NN$ which are not unions of units, and so cannot
be represented by any unit structure. On the other hand, one can always construct a rate matrix that
will implement any hypothesized unit structure over a set of coordinates, i.e., all unit structures
can actually exist, for some appropriate rate matrix. (At worst, one can do this by choosing a rate matrix in which each coordinate
evolves autonomously, i.e., a rate matrix that is a sum over all coordinates of independent rate matrices for 
each of those coordinates.) 
Not all collections of
units is a unit structure though; one can form a collection of units that contains two units $\oo, \oo'$, 
but not the unit $\oo \cap \oo'$, and so that collection won't be a unit structure.

For simplicity, from now on I assume that the unit structure doesn't change with $t$.
In addition, I define a conditional distribution for the ending joint state 
given an initial joint state, $p\left(x({t_f}) \,|\, x(t_i)\right)$, to be \textbf{consistent}
with a specified unit structure if there is some rate matrix that obeys that unit structure and that implements $p\left(x({t_f}) \,|\, x(t_i)\right)$.
The dynamics of any two units $\oo, \alpha \subset \oo$ 
must be compatible with one another, i.e., for all $p_{x_\oo}(t) = p_{x_\alpha, x_{\oo \setminus \alpha}}(t)$, 
%

\eq{
\sum_{x'_\alpha} K^{x'_\alpha}_{x_\alpha}(\alpha; t) p_{x'_\alpha}(t) &=  \sum_{x'_\alpha}
\sum_{x'_{\oo \setminus \alpha}} \sum_{x_{\oo \setminus \alpha}} 
			K^{x'_\alpha, x'_{\oo \setminus \alpha}}_{x_\alpha, x_{\oo \setminus \alpha}}(\oo; t)  p_{x'_\alpha, x'_{\oo \setminus \alpha}}(t)
\label{eq:consistency_condition}
}
i.e.,
\eq{
\sum_{x'_\alpha} K^{x'_\alpha}_{x_\alpha}(\alpha; t) \sum_{x'_{\oo \setminus \alpha}} p_{x'_\alpha, x'_{\oo \setminus \alpha}}(t) &= 
	 \sum_{x_{\oo \setminus \alpha}} \sum_{x'_\alpha} \sum_{x'_{\oo \setminus \alpha}}
			K^{x'_\alpha, x'_{\oo \setminus \alpha}}_{x_\alpha, x_{\oo \setminus \alpha}}(\oo; t)  p_{x'_\alpha, x'_{\oo \setminus \alpha}}(t)
}
In particular \cref{eq:consistency_condition} must hold for 
\eq{
p_{x'_\alpha, x'_{\oo\setminus\alpha}}(t) = \delta\left([x'_\alpha, x'_{\oo\setminus\alpha}], [x''_\alpha, x''_{\oo\setminus\alpha}]\right)
}
for any joint state $[x''_\alpha, x''_{\oo\setminus\alpha}]$. If we 
apply this requirement for all such delta function choices of $p_{x'_\alpha, x'_{\oo\setminus\alpha}}(t)$
and then relabel, we see that for all
$x_\alpha, x'_\alpha$, and $x'_{\oo \setminus \alpha}$,
\eq{
 K^{x'_\alpha}_{x_\alpha}(\alpha; t) &=  \sum_{x_{\oo \setminus \alpha}} 
			K^{x'_\alpha, x'_{\oo \setminus \alpha}}_{x_\alpha, x_{\oo \setminus \alpha}}(\oo; t)
\label{eq:consistency_rate_matrices}
}
(See App.\,B for more discussion of this result.) 
Conversely, if there is some rate matrix $K^{x'_\alpha}_{x_\alpha}(\alpha; t)$ such that \cref{eq:consistency_rate_matrices} holds
for all $x'_{\oo \setminus \alpha}$, then the two rate matrices in \cref{eq:consistency_rate_matrices} are 
compatible with each other, i.e., \cref{eq:consistency_condition} holds.
As an important special case of \cref{eq:consistency_rate_matrices}, if we take $\oo = \NN$ and as shorthand 
write $K(\NN; t)$ as just $K(t)$, we see that for any unit $\alpha$,
\eq{
 K^{x'_\alpha}_{x_\alpha}(\alpha; t) &=  \sum_{x_{-\alpha}} 
			K^{x'_\alpha, x'_{- \alpha}}_{x_\alpha, x_{-\alpha}}(t)
\label{eq:2.5}
}
independent of $x'_{-\alpha}$.

\begin{example}
Recall that an MPP is a set of co-evolving subsystems evolving according to a CTMC in which no transitions
are allowed in which more than two subsystems both change their states. Formally, in an MPP,
for all subsystems $i$, for all $x', x$, $K^{x'_i, x'_{-i}}_{x_i,x_{-i}}(t) = 0$ unless 
$x'_{-i} = x_{-i}$~\cite{horowitz_multipartite_2015,kolchinsky_wolpert_2020_thermo_under_constraints}. Equivalently, 
for every subsystem $i$ in an MPP, there is an associated rate matrix $K^{x'}_{x}(i; t)$ 
which is zero if $x'_{-i} \ne x_{-i}$ such that the global rate $K$ matrix can be written as
\eq{
K^{x'}_x(t) = \sum_{i \in \NN} K^{x'_i, x'_{-i}}_{x_i, x_{-i}}(i ; t) 
\label{eq:mpp}
}
and where for every unit $\oo$ containing subsystem $i$, the rate matrix terms
$K^{x'_\oo, x'_{-\oo}}_{x_\oo, x'_{-\oo}}(i; t)$ are independent of $x'_{-\oo}$.

The
units in an MPP are sets of subsystems whose joint evolution is independent of the other subsystems. 
\cref{eq:consistency_condition} 
often holds (and therefore so does \cref{eq:consistency_rate_matrices}) in an MPP. At the other extreme
from multipartite processes, \cref{eq:consistency_rate_matrices} also holds for some rate matrices $K(t)$ which only allow state transitions
in which \underline{all} subsystems change, i.e., rate matrices $K(t)$ such that $K^{x'}_x(t) = 0$ for any $x, x'$
where there are two subsystems, $j, k$ such that both $x_j \ne x'_j$ and  $x_k = x'_k$. This is illustrated in App.\,B.
\label{ex:MPP}
\end{example}

It will often be convenient to re-express a unit structure as a directed graph.
Define the \textbf{dependency graph} $\Gamma_{\NN^*} = ({\NN^*}, E)$ by the rule that there
is an edge $e \in E$ from node $\oo \in \NN^*$ to node $\oo' \in \NN^*$ iff both:
$\oo' \subseteq \oo$, and there is no intervening unit $\oo''$ such that $\oo' \subseteq \oo'' \subseteq \oo$.
(Note that $\Gamma_{\NN^*}$ is a directed graph, which allows us to use standard
graph theory terminology.) 
In a unit structure ${\NN^*}$ where $\NN \in {\NN^*}$ the dependency graph
has a single root, but if  $\NN \not \in {\NN^*}$, then the dependency graph has
multiple roots. 

I will abuse notation and sometimes treat a unit $\oo$ as a set of coordinates while at other times
I treat it as a single node in $\Gamma_{\NN^*}$. I  write the set of parents of any node $\oo \in \Gamma_{\NN^*}$ as $\pa(\oo)$, 
and the set of its descendants as $\desc(\oo)$, with $\family(\oo) := \oo \cup \desc(\oo)$, the \textbf{family} of node $\oo$.
The maximal number of nodes in any directed path that starts at $\oo$ is
the \textbf{height} of $\oo$. So any unit $\oo$ which has
no sub-units contained in it is a leaf node of $\Gamma_{\NN^*}$, with height $1$. 
(The maximal height of all nodes in $\Gamma_{\NN^*}$ is simply called ``the height of $\NN^*$''.)
I write $\Gamma_{\NN^*}^R$ for the set of root nodes in $\Gamma_{\NN^*}$.
As an example, the dependency graph of \cref{fig:1} has two root nodes, $\oo$ and $\alpha$, and one
leaf node, $\oo'$, which is their common child. The height of the graph is $2$. 

There are several additional, technical conditions that I will impose on the unit structure, in order
to simplify the algebra in the proofs of the results in \cref{sec:stronger_bound}. (These conditions can
be ignored if the reader is only interested in understanding the results, not the details of their proofs.)
\begin{enumerate}

\item I require that the unit structure is rich enough that
if a joint state transition can occur that simultaneously changes the state of all coordinates in a set $\alpha$, then
there is some unit $\oo \in \NN^*$ that contains $\alpha$.\footnote{Formally, if 
there is some $x', x$ such that $K^{x'}_x(t) \ne 0$ while $x'_i \ne x_i$ for all $i \in \alpha$,
then there is some unit $\oo \in \NN^*$ where $\alpha \subseteq \oo$. } I call such a unit structure \textbf{flush}.

\item A unit $\oo$ is \textbf{vacuous} if every one of its coordinates are in at least 
one subunit $\oo' \subseteq \oo$.
I assume that no unit in any unit structure we are considering is
vacuous.\footnote{For a vacuous unit $\oo$, 
the dynamics of $x_\oo$ is fully specified by the rate matrices
of the subunits $\oo' \subseteq \oo$, and so to specify a rate matrix for $\oo$ would be redundant.}

\item I say that two units $\oo, \oo' \subset \oo$ are \textbf{equivalent} at time $t$ if
for all $x'$ where $p_{x'}(t) \ne 0$, for all $x$ such that $x'_{\oo \setminus \oo'} \ne x_{\oo \setminus \oo'}$,
$K^{x'}_x(\oo; t) = 0$. I require that $\NN^*$ does not contain any two equivalent units. This means that for any two units
$\oo, \oo' \subset \oo$ in the unit structure, there must be transitions $x' \rightarrow x$ that can occur in which
some coordinate $i \in \oo \setminus \oo'$ changes its value.
\end{enumerate}
Any CTMC can be represented with at least one unit structure meeting these three conditions (e.g., the unit structure
that consists just of $\{\NN\}$). 

To connect these considerations to the theorems of stochastic thermodynamics, 
from now on I suppose we can model the system as though there are a total of $R$ thermodynamic reservoirs attached to the 
system~\cite{van2015ensemble,seifert2012stochastic}. I suppose further that
each reservoir $v \in \{1, \ldots, R\}$ generates fluctuations of the joint state of
an associated set of coordinates $\PP(v) \subseteq \NN$, without any such direct effect on the other coordinates.
(For example, $v$ may be able to do this by being directly physically coupled to the coordinates in $\PP(v)$
and no others, via an implicit interaction Hamiltonian.) As is standard in stochastic thermodynamics, I suppose
that if only one particular reservoir $v$ were attached to the system, then the resultant dynamics over $\PP(v)$
would be a CTMC. $\PP(v)$ is called the \textbf{puppet set} of reservoir $v$, with its elements called the \textbf{puppets} of $v$. 
The collection of all $R$ puppet sets covers $\NN$. 


\begin{example}
Return to the example of an MPP, where we identify each subsystem 
with a separate coordinate. Each subsystem has its own unique set of reservoirs,
which jointly causes the fluctuations in its state.
In other words, the puppet set of each reservoir is a singleton, the associated
subsystem of that reservoir, and each subsystem is the puppet set of
at least one reservoir. On the other hand, in general, the rate matrices of each subsystem $i$
will depend on the states of other subsystems besides $i$. When that is the case, the leader
set of the reservoirs of each subsystem $i$ will not be a singleton.
\end{example}

\noindent For simplicity, from now on I assume that neither the number of reservoirs
nor the associated maps $\PP(.)$ and $\LL(.)$ changes with time $t$.

I write $\LL(v) \supseteq \PP(v)$ for a set of coordinates
whose associated value directly affects how the coupling with reservoir $v$ affects the dynamics of $x_{\PP(v)}$.
I call this the \textbf{leader set} of $\PP(v)$, or sometimes the leader set of $v$.\footnote{Physically, $\LL(v)$ will often
reflect an interaction Hamiltonian coupling the coordinates in $\LL(v)$ without any back-action of the value
of $x_{\PP(v)}$ onto the values of $x_{\LL(v) \setminus \PP(v)}$. This isn't necessary for the analysis below though.}
I write $L(v; t)$ for the associated rate matrix over $X$ induced by the coupling of the system to reservoir $v$.
So $L(v; t)$ affects of dynamics of $x_{\PP(v)}$, but leaves the other coordinates unchanged. 
I write this as
\eq{
L^{x'}_x(v; t) 		&= 	L^{x'_{\LL(v)}}_{x_{\PP(v)}, x_{\LL(v) \setminus \PP(v)}}(v; t) \delta^{x'_{_{-\PP(v)}}}_{x_{_{-\PP(v)}}} 
\label{eq:2.8a}
}
where $L^{x'_{\LL(v)}}_{x_{\PP(v)}, x_{\LL(v) \setminus \PP(v)}}(v; t)$ is a proper stochastic rate matrix that
equals $0$ if $x_{\LL(v) \setminus \PP(v)} \ne x'_{\LL(v) \setminus \PP(v)}$. 
As in conventional stochastic thermodynamics, the global rate matrix at time $t$ is
the sum over all reservoirs of the rate matrices of those reservoirs,
\eq{
K^{x'}_x(t) 
	&=  \sum_v 
L^{x'_{\LL(v)}}_{x_{\PP(v)}, x_{\LL(v) \setminus \PP(v)}}(v; t) \delta^{x'_{_{-\PP(v)}}}_{x_{_{-\PP(v)}}} 
\label{eq:2.10}
}
I refer to any system evolving according to \cref{eq:2.10} for some associated set of puppet sets
leader sets, and matrices $L^{x'}_x(v; t)$ as a \textbf{composite system}.

In the rest of this section I introduce some notation that will be helpful in analyzing composite systems.
First, as shorthand I will sometimes 
rewrite \cref{eq:2.8a} as
\eq{
L^{x'}_x(v; t)	&= 	L^{x'_{\LL(v)}}_{x_{\PP(v)}}(v; t) \delta^{x'_{-\PP(v)}}_{x_{-\PP(v)}}
\label{eq:2.8}
}
where $L^{x'_{\LL(v)}}_{x_{\PP(v)}}(v; t)$ is a ``rate matrix'' in that all of its entries for $x'_{\PP(v)} \ne x_{\PP(v)}$
are non-negative, and 
\eq{
\sum_{x_{\PP(v)}} L^{x'_{\LL(v)}}_{x_{\PP(v)}}(v; t) = 0
}
See~\cref{fig:1} above and~\cref{ex:3} below.

In general, any given coordinate $i$ may be in more than one reservoir's leader set and in more
than one reservoir's puppet set. Accordingly, I extend the definitions above by writing
\eq{
\LL(i) &:= \bigcup_{v : i \in \PP(v)} \LL(v) \\
\LL(A) & := \bigcup_{i \in A} \LL(i)
}
where $A$ is an any subset of $\NN$. So $\LL(i)$ is the set of all coordinates whose state
can directly affect the dynamics of coordinate $i$, via arguments of a rate matrix, and similarly for $\LL(A)$. Along
the same lines, I define
\eq{
\PP(i) &:= \bigcup_{v : i \in \PP(v)} \PP(v) \\
\PP(A) & := \bigcup_{i \in A} \PP(i)
}
So $\PP(A)$ is the set of all coordinates, inside or outside of $A$, whose dynamics is governed jointly with that of any coordinate in $A$.
$\LL(v) \subseteq \LL(\PP(v))$, since there can be coordinates $i \in \PP(v)$ whose
dynamics is affected by other reservoirs in addition to $v$. Note as well that for any set $A$, $A \subseteq \PP(A) \subseteq \LL(A)$.
So in particular,  if any two different units have nonempty intersection,
then since that intersection must also be a unit, the leader sets of all the coordinates in
that intersection must lie within that intersection. 
In addition, the inverses of these set-valued functions are well-defined. In particular, for any set of coordinates $A$, 
$\PP^{-1}(A)$ is the set of all reservoirs $v$ such that $i \in \PP(v)$ for some $i \in A$. 

It will be convenient to introduce the shorthand that for any subset $A \subseteq \NN$, $\nu(A)$ is the set of all reservoirs $v$ such that
$\PP(v) \cap A \ne \varnothing$. So $\nu(A)$ is the set of all reservoirs who affect the dynamics of any of the coordinates 
in $A$. In \cref{app:prop_1}  it is shown that \cref{eq:2.10} implies the following intuitive result:
\begin{proposition}
\label{prop:1a}
For any unit $\oo$,
\eq{
K^{x'_\oo}_{x_\oo}(t) &=  \sum_{v \in \nu(\oo)} 
		\widehat{L}^{x'_{\LL(v) \cap \oo}}_{x_{\PP(v) \cap \oo}}(v, \oo;t)
		\delta^{x'_{_{\oo \setminus \PP(v)}}}_{x_{_{\oo \setminus \PP(v)}}}  \nonumber
}
where $\widehat{L}^{x'_{\LL(v) \cap \oo}}_{x_{\PP(v) \cap \oo}}(v, \oo;t)$ is a properly normalized rate matrix
over $x_{\PP(v) \cap \oo}$ and is independent of $x'_{_{\oo \setminus \PP(v)}}$.
\end{proposition}
\begin{example}
As a simple illustration of \cref{prop:1a}, consider any MPP where each subsystem is controlled by one reservoir, which
controls no other subsystems. To reduce notation, consider the case where the unit $\oo$ is 
all of $\NN$. In this case the sum over $v \in \nu(\oo)$ runs over all subsystems $i$ in unit $\oo$, and each
$ \widehat{L}^{x'_{\LL(v) \cap \oo}}_{x_{\PP(v) \cap \oo}}(v, \oo;t)$ is the rate matrix of the subsystem $i$ associated with reservoir $v$.
So \cref{prop:1a} reduces to \cref{eq:mpp} in \cref{ex:MPP}, with 
each term $ \widehat{L}^{x'_{\LL(v) \cap \oo}}_{x_{\PP(v) \cap \oo}}(v, \oo;t) 	\delta^{x'_{_{\oo \setminus \PP(v)}}}_{x_{_{\oo \setminus \PP(v)}}} $
in \cref{prop:1a} re-expressed as $K^{x'_i, x'_{-i}}_{x_i, x_{-i}}(i ; t)$.
\end{example}

\cref{prop:1a} means that as far as any single unit $\oo$ is concerned, we can replace each reservoir $v \in \nu(\oo)$, which has leader
set $\LL(v)$ and puppet set $\PP(v)$, with a reservoir which has leader set $\LL(v) \cap \oo  \subseteq \oo$ and puppet set $\PP(v) \cap \oo \subseteq \oo$.
For simplicity I assume in the analysis below that 
we have chosen a unit structure where all such replacements have been made. Formally, 
without loss of generality, I restrict attention to unit structures that only contains 
units $\oo$ with the property that for all reservoirs $v \in \nu(\oo)$, $\LL(v) \subseteq \oo$. I call such a unit structure \textbf{tight}. (Note that
there is always at least one unit structure with this property, namely the unit structure with a single element, the unit $\NN$.)

We can tighten \cref{prop:1a} under our assumption of a tight unit structure. The following result is proven in \cref{app:proof_prop:1b}:
\begin{proposition}
\label{prop:1b}
For any unit $\oo$ in a tight unit structure,
\eq{
K^{x'_\oo}_{x_\oo}(\oo; t) &= \sum_{v \in \nu(\oo)} L^{x'_{\oo}}_{x_{\oo}}(v; t) \nonumber
}
\end{proposition}
 
For any unit $\oo$ in a tight unit structure, $\LL(\oo) = \oo$.\footnote{To see this, first use the
fact that for every unit $\oo$ in any unit structure, whether that unit structure is tight or not, 
$\oo \subseteq \LL(\oo)$. Then note that if $\oo \subset \LL(\oo)$, at least one reservoir in $\LL(\oo)$ would violate the
requirement that the unit structure be tight.}
Since $A \subseteq \PP(A) \subseteq \LL(A)$ for all sets $A$, it then follows that
 $\PP(\oo) = \oo$ for any unit $\oo$ in a tight unit structure. 
So loosely speaking, no reservoir is allowed to ``straddle'' coordinates lying both within
a unit and outside a unit, if we restrict attention to tight unit structures. In addition, $\PP(-\oo) = -\oo$
in a tight unit structure, even though $-\omega = \NN \setminus \oo$ is not a unit in
general.\footnote{To see this, first note that $-\oo\subseteq \PP(-\oo)$, since any set is contained in the associated puppet set. In addition, if there were coordinates 
$i \in \PP(-\oo)$ that lay inside $\oo$ for some reservoir $v$, then there would be a coordinate $j \in -\oo$ such that $\PP(v)$
contains both $i$ and $j$. That in turn would mean that $j \in \PP(\oo)$ and therefore $j \in \LL(\oo)$. However, that violates the assumption
that the unit structure is tight. Therefore $\PP(-\oo) \subseteq -\oo$. Combining establishes the claim. }

\section{Thermodynamics of composite systems}


Following conventional stochastic thermodynamics, I identify the (expected) \textbf{global } EF rate at time $t$ as
\eq{
\langle \dot{Q}(t) \rangle &= \sum_{v, x', x} L^{x'}_x(v; t) p_{x'}(t) \ln \dfrac{L^{x'}_{x}(v; t)}{L^{x}_{x'}(v; t)} \\
	&= \sum_{v} \sum_{x'_{\LL(v)}} \sum_{x_{\LL(v)}} 
	L^{x'_{\LL(v)}}_{x_{\PP(v)}}(v; t)  p_{x'_{\LL(v)}}(t) 
			\ln \dfrac{L^{x'_{\LL(v)}}_{x_{\PP(v)}}(v; t) } {L^{x_{\LL(v)}}_{x'_{\PP(v)}}(v; t)} 
\label{eq:3.6}
}
where the second equality is established in
App.\,C.

The results below do not require LDB. However, if all reservoirs are purely thermal, with no associated particle
exchange, and if LDB applies, then we can interpret 
the EF rate as (temperature-normalized) heat flow between the system and its 
reservoirs.\footnote{Note that for convenience in the definition of ``in-ex information'' below, 
I adopt the convention that $\langle \dot{Q}(t) \rangle$ is the expected EF 
rate \textit{out of} the system into the reservoirs, and so is the negative of the integrand in \cref{eq:2}.}

Similarly, the (expected) \textbf{global } EP rate at time $t$ is the difference between the time-derivative of the global entropy
and the expected EF rate. Expanding, we can write that difference as
\eq{
\langle \dot{\sigma}(t) \rangle &= \sum_{v, x', x} L^{x'}_x(v; t) p_{x'}(t) \ln \dfrac{L^{x}_{x'}(v; t)  p_{x}(t)}{L^{x'}_x(v; t)  p_{x'}(t)} \\
		&= \sum_{v} \sum_{x'_{\LL(v)}, \,x_{\LL(v)}} L^{x'_{\LL(v)}}_{x_{\PP(v)}}(v; t) p_{x'_{\LL(v)}}(t) 
			\ln \dfrac{L^{x'_{\LL(v)}}_{x_{\PP(v)}}(v; t) p_{x'_{\LL(v)}}(t)} {L^{x_{\LL(v)}}_{x'_{\PP(v)}}(v; t) p_{x'_{\LL(v)}}(t)}
\label{eq:2.14a}
}
\cref{eq:3.6,eq:2.14a} formally establish that the thermodynamics associated with each reservoir $v$
doesn't involve any coordinates outside of $\LL(v)$, just as one would expect.

\begin{example}
In an MPP with a single reservoir per system, each coordinate $i$ is a ``subsystem''; $R = N$;
there is a bijection between the set of reservoirs and the set of subsystems; and
for every reservoir / subsystem $i$, $\PP(i) = \{i\}$. 
So a unit $\oo$ is any set of subsystems such that for all $i \in \oo$, $\LL(i) \subseteq \oo$.
In addition, \cref{eq:2.14a} reduces to
\eq{
\langle \dot{\sigma}(t) \rangle 
		&= \sum_{i} \sum_{x'_{\LL(i)}, \,x_{\LL(i)}} L^{x'_{\LL(i)}}_{x_{i}}(v; t) p_{x'_{\LL(i)}}(t) 
			\ln \dfrac{L^{x'_{\LL(i)}}_{x_{i}}(i; t) p_{x'_{\LL(i)}}(t)} {L^{x_{\LL(i)}}_{x'_{i}}(i; t) p_{x_{\LL(i)}}(t)}
}
See~\cite{horowitz_multipartite_2015,wolpert2020minimal,wolpert2020fluctuation,wolpert.thermo.bayes.nets.2020,hartich_sensory_2016,Brittain_2017}
and~\cref{fig:1}.
\label{ex:3}
\end{example}

Following the same convention as for global EF rate, 
I define the (expected) \textbf{local} EF rate of any unit $\oo \subseteq \NN$ at time $t$ as
the entropy flow rate into the associated reservoirs:
\eq{
\langle\dot{Q}^{\oo} (t)\rangle &= 
	\sum_{v \in \nu(\oo)} \sum_{x', x} L^{x'}_x(v; t) p_{x'}(t) \ln \dfrac{L^{x'}_{x}(v; t)}{L^{x}_{x'}(v; t)} \\
	&= 
	\sum_{x'_\oo, x_\oo, v \in \nu(\oo)} L^{x'_\oo}_{x_\oo}(v; t) p_{x'_\oo}(t)
					 \ln \dfrac{L^{x'_\oo}_{x_\oo}(v; t)}{L^{x_\oo}_{x'_\oo}(v; t)}
\label{eq:4}
}
Since no reservoir's puppet set can include both coordinates inside a unit $\oo$ and coordinates outside of $\oo$,
for any two units $\oo, \oo'$ where $\oo \cap \oo' = \varnothing$,
\eq{
\langle\dot{Q}^{\oo \cup \oo'} (t)\rangle &= \sum_{x'_\oo, x_\oo, v \in \nu(\oo \cup \oo')} L^{x'_\oo}_{x_\oo}(v; t) p_{x'_\oo}(t)
					 \ln \dfrac{L^{x'_\oo}_{x_\oo}(v; t)}{L^{x_\oo}_{x'_\oo}(v; t)}  \\
	&= \sum_{x'_\oo, x_\oo}  \left[\sum_{v \in \nu(\oo)} L^{x'_\oo}_{x_\oo}(v; t) p_{x'_\oo}(t)
					 \ln \dfrac{L^{x'_\oo}_{x_\oo}(v; t)}{L^{x_\oo}_{x'_\oo}(v; t)}   +  \sum_{v \in \nu(\oo')}L^{x'_\oo}_{x_\oo}(v; t) p_{x'_\oo}(t)
					 \ln \dfrac{L^{x'_\oo}_{x_\oo}(v; t)}{L^{x_\oo}_{x'_\oo}(v; t)} \right]  \\
   &= \langle\dot{Q}^{\oo} (t)\rangle + \langle\dot{Q}^{\oo'} (t)\rangle
\label{eq:3.8}
}

So viewed as a function from the set of all units to reals, $\langle \dot{Q}^{\oo} (t)\rangle$ obeys 
the countable additivity axiom of a signed measure over  $\SSSS(\NN^*)$, the sigma algebra generated by the
units in $\NN^*$.
%
This allows us to extend the definition of local EF rate to the sigma algebra $\SSSS(\NN^*)$
by using the set of values $\{\langle\dot{Q}^{\oo} (t)\rangle : \oo \in \NN^*\}$ to generate an entire signed measure.
So for example, for every pair of units $\oo, \oo' \subset \oo$ in $\NN^*$, even if $\oo \setminus \oo' \not\in \NN^*$,
\eq{
\langle\dot{Q}^{\oo \setminus \oo'} (t)\rangle := \langle\dot{Q}^{\oo} (t)\rangle - \langle\dot{Q}^{\oo'} (t)\rangle
}

Recall that the dynamics of any unit is given by a self-contained CTMC, independent of the state of
any coordinate outside of that unit. Accordingly,
the EP rate of a unit is the sum of the derivative of the entropy of the distribution
of the joint state of that unit and the EF rate into the reservoirs of that unit. Using
\cref{app:proof_prop:1b} to evaluate that entropy derivative and \cref{eq:4} to evaluate that EF rate,
we see that the (expected) \textbf{local} EP rate of $\oo$ at time $t$ is
\eq{
\label{eq:15}
\langle\dot{\sigma}^{\oo} (t) \rangle &= \dfrac{d S^\oo(t)}{dt} + \langle\dot{Q}^\oo (t)\rangle    \\
	&= \sum_{x'_\oo,x_\oo,v \in \nu(\oo)}  L^{x'_\oo}_{x_\oo}(v; t) p_{x'_\oo}(t)
					 \ln \left[\dfrac{L^{x'_\oo}_{x_\oo}(v; t)  p_{x'_\oo}(t)}  {L^{x_\oo}_{x'_\oo}(v; t)  p_{x_\oo}(t)} \right] 
\label{eq:5}
}
Accordingly I sometimes write the {global} EP rate given in \cref{eq:2.14a} as $\dSt$.
%
For any unit $\oo$, $\dSot \ge 0$, since
$\dSot$ has the usual form of an EP rate of a single system. 
(See~\cite{kolchinsky_wolpert_2020_thermo_under_constraints} for a discussion of the relation between local EP rates
and similar quantities discussed in~\cite{shiraishi2015fluctuation,horowitz2014thermodynamics,horowitz_multipartite_2015}.)

Write the local EP generated by a unit $\oo$ during the process as
\eq{
\sigma^\oo := \int_{t_i}^{t_f} dt \, \langle \dot{\sigma}^\oo \rangle
}
and similarly write $\sigma^\NN$ for the global EP. (To minimize notation,
I adopt the convention that angle brackets are implicit for time-extended thermodynamic quantities, as opposed to rates.)
In App.\,D, \cref{eq:consistency_rate_matrices} and the log sum inequality~\cite{cover_elements_2012} are used to prove that for any two units $\oo, \alpha \subset \oo$, not
necessarily part of a unit structure, $\dSot \ge \langle\dot{\sigma}^{\alpha} (t) \rangle$ at all times $t$.
Therefore 
\eq{
\sigma^\oo \ge \sigma^{\alpha}
\label{eq:13}
}
%
In particular it is shown in~\cite{wolpert_thermo_comp_review_2019,kolchinsky_wolpert_2020_thermo_under_constraints} 
that in the special case where there is a
set of units  $\{\alpha_j\}$ who have no overlap with another, for any unit $\oo \supset \cup_j \alpha_j$, 
\eq{
\sigma^\oo \ge \sum_j \sigma^{\alpha_j}
\label{eq:13a}
}
(See also \cref{eq:24} below.)

%
%

Let ${\NN^*} = \{\oo_j : j = 1, 2, \ldots, n\}$ be a unit structure. For simplicity, from now on I assume that $\NN \not \in \NN^*$.
Suppose we have a set of real numbers, $f$, which are indexed by the units ${\NN^*}$. It  will be convenient
to use the associated shorthand,
\eq{
\SSS_{\oo \in {\NN^*}} f^{\oo} &:= \sum_{j = 1}^n f^{\oo_j} - \sum_{1 \le j < j' \le n} f^{\oo_j \cap \oo_{j'}}
+ 	\sum_{1 \le j < j' < j'' \le n} f^{\oo_j \cap \oo_{j'} \cap \oo_{j''}} - \ldots
\label{eq:in_ex_gen}
}
(Note that the precise assignment of integer indices to the units in $\NN^*$ is irrelevant.)
This quantity is called the \textbf{inclusion-exclusion sum} (or just ``in-ex sum'' for short) of $f$ for the unit structure $\NN^*$.

Next, define the time-$t$ \textbf{in-ex information} as
\eq{
&\II^{\NN^*} \;:=\; \left[\widehat{\sum_{\oo \in {\NN^*}}}   S^\oo\right] - S^\NN 
\;=\; - S^\NN + \sum_{j = 1}^n S^{\oo_j} - \sum_{1 \le j < j' \le n} S^{\oo_j \cap \oo_{j'}} + \ldots
\label{eq:in_ex_info}
}
where all the terms in the sums on the RHS are marginal entropies over the (distributions over the coordinates in) the indicated units.
As an example, if $\NN^*$ consists of two units, 
$\oo_1, \oo_2$, with no intersection, then the expected in-ex information at time $t$ is just
the mutual information between those units at that time.
More generally, if there an arbitrary number of units in ${\NN^*}$
but none of them overlap, then the expected in-ex information is what is called the ``multi-information'', or
``total correlation'', among those units~\cite{mcgill1954multivariate,ting1962amount,kolchinsky_wolpert_2020_thermo_under_constraints,ting1962amount}.

In App.\,E, Rota's extension of the inclusion-exclusion
principle~\cite{stanley2011enumerative} is used to show that in any composite unit structure,
\eq{
\langle\dot{Q}^\NN (t)\rangle &= \SSS_{\oo \in {\NN^*}} \langle\dot{Q}^\oo (t)\rangle
\label{eq:11aa}
}
This implies that the global EP rate is 
\eq{
\dSt	&\;=\; 
\dfrac{d S^\NN(t)}{dt} + \langle\dot{Q}^\NN (t)\rangle  
\;=\; 
-\dfrac{d}{dt} \II^{\NN^*}(t) + \SSS_{\oo \in {\NN^*}} \langle\dot{\sigma}^\oo (t)\rangle
\label{eq:11a}
}
This is the first major result of this paper.\footnote{A related result 
was derived in~\cite{kolchinsky_wolpert_2020_thermo_under_constraints}, applicable to the special case where the
composite system is a multipartite process, and it
obeys a form of LDB. The derivation here is the first that shows the result applies more generally, even when
those two conditions are violated.}
Integrating \cref{eq:11a} from the start to the end of a process gives 
\eq{
\sigma^\NN &= \SSS_{\oo \in {\NN^*}}\sigma^\oo - \Delta \II^{\NN^*} 
\label{eq:12}
}

As an example of this result, suppose that we have two physically separated subsystems undergoing an MPP, and that
subsystem $2$ never changes its state, while subsystem $1$ executes a map 
from $p_{x_1}(t_i)$ to $p_{x_1}(t_f)$, independent of the state of $x_2$. 
Note that if the rate matrix of subsystem $1$ depends on the state of subsystem $2$, i.e., 
subsystem $1$ observes subsystem $2$ as it evolves, then there is only one unit rather than two. Accordingly,
\cref{eq:11a} tells us that the global EP rate can
depend on this property of whether subsystem $1$ observes the state of subsystem $2$ as subsystem $1$ evolves, 
even though the conditional distribution of subsystem $1$'s final state given its initial state,
$p(x_1(t_f) | x_i(t_i), x_2(t_i))$, is independent of the state of subsystem $2$. 
In general, this effect of the unit structure on the EP will occur whenever the two subsystems are initially statistically coupled.  
See 
App.\,F 
for a discussion.

\cref{eq:12} applies to any unit structure.
In addition, for any unit structure $\MM^*$ over a set of coordinates $\MM \subset \oo$, 
\cref{eq:13} and the fact that the union of a set of units is itself a unit means that $\sigma^{\oo} - \sigma^{\MM} \ge 0$. 
Therefore using \cref{eq:12} to expand $\sigma^\MM$ gives
\eq{
\sigma^\oo - \SSS_{\oo' \in {\MM^*}}\sigma^{\oo'} &\ge  -\Delta \II^{\MM^*}
\label{eq:21}
}
\cref{eq:21} holds even if $\MM^* \subset \NN^*$, and at the other extreme,
even if no unit in $\MM^*$ is also in $\NN^*$.

Finally, as an aside, I note that if local detailed balance holds for all reservoirs $v$ with puppet set inside a unit, then all the usual fluctuation theorems~\cite{seifert2012stochastic},  thermodynamic uncertainty
relations~\cite{gingrich_horowitz_finite_time_TUR_2017,horowitz_gingrich_nature_TURs_2019,liu2020thermodynamic},
first-passage time bounds~\cite{gingrich2017fundamental}, bounds on stopping times~\cite{neri2017statistics}, etc.,
apply to the thermodynamics of that unit. See~\cite{wolpert_mix_match_encyclopedia_2022} for extensive
analysis of the implications of this for the special case of composite systems that are MPPs.


\section{Strengthened second law for composite systems}
\label{sec:stronger_bound}


In general, to evaluate the in-ex sum of local EPs on the RHS of \cref{eq:12} requires detailed knowledge of the precise rate matrices
during the process. However, following Landauer, the goal in this paper is to derive bounds that are independent
of those details, depending only on the starting distribution and the conditional distribution of the final state given
the initial state. 
%
One might hope that one could achieve this goal simply by setting all local EPs to $0$ in \cref{eq:12}, giving 
\eq{
\sigma^\NN &\ge  -\Delta \II^{\NN^*} \\
	&:= \BBB_{\NN^*}
\label{eq:height_2}
}

Unfortunately,  in general it is impossible to have the local EPs of all units $= 0$ in an arbitrary unit structure,
even if one uses a quasistatically slow process. Indeed, the unit structure itself, independent of any other properties of
the rate matrix, may mean that it is impossible to have all local EPs $=0$.\footnote{As an
example, if one of the units $\oo$ has a set of units within it, and if each of those units \textit{within} $\oo$
has zero local EP but that set of units within $\oo$ has nonzero
change in its in-ex information during the process, then the local EP of $\oo$ as a whole is nonzero.}

This might seem to imply that we cannot lower-bound the EP as $\BBB_{\NN^*}$.
However, recall that in general there are many different unit structures that all apply to the
same CTMC. We are free to choose among those unit structures.
And as it turns out, no matter what the CTMC is, we can always choose the unit structure in a way that
guarantees that \cref{eq:height_2} does in fact hold. 

I prove this result in several steps. First, in  App.\,F and App.\,G, I 
derive a set of lower bounds on EP that always apply, no matter what the
unit structure. These lower bounds are summarized in Prop.\;F.1, and are my second main result. 
These bounds are not in the form of \cref{eq:height_2} though; while important in their
own right, they do not yet achieve our goal. 

On the other hand, in general we can represent {any} CTMC with a unit structure of height $2$. (For example, we can
do that by combining all coordinates that are not members of a root node of
$\Gamma_{\NN^*}$, into one, overarching unit.) In App.\,F I derive
a corollary of Prop.\;F.1, telling us that  \cref{eq:height_2} holds for any such unit structure of height 2.
This is my third main result.\footnote{It is also shown explicitly in that appendix there are unit
structures of arbitrary height that obey \cref{eq:height_2}, in addition to unit structures that violate it.
See Prop.\;F.2.}

Due to this third result, we can always choose the unit structure $\NN^*$ so that the global EP is bounded by~\cref{eq:height_2}.
Unfortunately, as illustrated below, there are some unit structures $\NN^*$ of height $2$ where the bound on the RHS of
\cref{eq:height_2}  is negative
for an appropriate initial distribution $p_{t_i}(x)$ and conditional distribution $p(x({t_f}) \,|\, x(t_i))$ consistent with
$\NN^*$. In such cases, \cref{eq:height_2} does not provide a stronger bound on EP than the conventional second law.
This is not as much of a problem as one might fear though. For \textit{every} unit structure $\NN^*$, 
there are initial distributions $p_{t_i}(x)$ and conditional distributions $p(x({t_f}) \,|\, x(t_i))$ that are consistent with
$\NN^*$ where the RHS of \cref{eq:height_2} is non-negative, so that the bound in \cref{eq:height_2}
is at least as strong as the conventional second law.
This is my third and final main result. (This result is presented in Prop.\,F.2, and is also proven in  App.\,F, based on results in App.\,I.)




%

\section{Thermodynamics of feedback control for composite systems}

We can use \cref{eq:height_2} to extend previous work on the thermodynamics of feedback 
control~\cite{sagawa2009minimal,parrondo2015thermodynamics, kolchinsky_wolpert_2020_thermo_under_constraints} to account
for a known set of dependency constraints of the system being controlled.
Suppose we have a composite system with some associated unit structure $\NN^*$ and
some desired initial and final joint distributions over the states 
of the system, $p^\dagger_{t_i}(x)$ and $ p^\dagger_{t_f}(x)$, respectively. Suppose we also
have a feedback controller, $\CC$, 
whose state space $C$ has values $c$. Before the system starts to evolve, 
the controller observes the initial state of the system through a noisy channel, 
$p(c | x)$. This observation does not affect that initial system state, i.e., there is no back-action.
So the initial joint distribution immediately after the observation is 
\eq{
p_{t_i}(c, x) =  p(c | x) p^\dagger_{t_i}(x)
\label{eq:p_ti}
}
As is standard in the literature of the thermodynamics of feedback 
control, we do not consider the thermodynamics of this measurement process.
Note that $p_{t_i}(x) =  p^\dagger_{t_i}(x)$.

After the measurement, the controller's state, $c$, does not change. However, the system
can observe $c$ as it evolves (or as it's more usually phrased, the state of $c$ serves
to ``control'' the state of the system). The result is a new final distribution, 
\eq{
p_{t_f}(c, x) = \sum_{x'} p(c | x') p^\dagger_{t_i}(x') p(x_{t_f} | x_{t_i} = x', c)
\label{eq:p_tf}
}
where we abuse notation and write $p(x_{t_f} | x_{t_i}, c)$ for the distribution
over final states of the system conditioned on the initial state being $x'$ and the feedback process state being $c$.
For simplicity we parallel the conventional analysis in the literature
and require that the marginal final distribution obeys $\sum_c p_{t_f}(c, x) = p^\dagger_{t_f}(x)$. 
In order to analyze the thermodynamics of feedback control, one must define a Hamiltonian over the states of the system, 
so that one can define the work on / from the system. Following
convention, I assume the Hamiltonian is uniform at both $t_i$ and $t_f$, and assume it is related to the
global rate matrix via LDB.

Let $\NN^*$ be some unit structure  with height less than $3$ representing the original system, without the feedback apparatus.
Using \cref{eq:height_2}, the EP without the feedback apparatus is lower-bounded by
\eq{
\sigma^\NN \ge \BBB_{\NN^*} = \II^{\NN^*}(p^\dagger_{t_i}(X)) - \II^{\NN^*}(p^\dagger_f(X))
}
By coupling that original system to the feedback apparatus we construct a new system, ${\MM}$, which comprises the original system
together with an extra subsystem (the feedback apparatus)
and new dependencies of the original coordinates of the system on the state of that new subsystem. There are
many possible unit structures, $\MM^*$, over this new joint system-feedback-apparatus. For simplicity, 
exploit the fact that $\CC$ evolves independently of the other coordinates in the system (by not evolving
at all) to construct $\MM^*$ directly from $\NN^*$, 
by replacing each unit $\oo \in \NN^*$ with a new unit, $\oo'(\oo) := \oo \cup {\CC}$. So $\MM^*$ and $\NN^*$
contain the same number of units, with each unit in $\MM^*$ containing the subsystem $\CC$,
and $X_{\oo'(\oo)} = C \times X_\oo$.
(It doesn't
matter if we add an additional unit to $\MM^*$, containing just $\CC$ itself.) This gives a new
lower bound on the EP, $\BBB_{\MM^*}$.
In App.\,J, it is shown that the difference between the lower bound on EP in the new, feedback scenario,
and the lower bound on EP in the original, no-feedback scenario, is
\eq{
\BBB_{\MM^*} - \BBB_{\NN^*} = \Delta \left[\SSS_{\oo \in {\NN^*}}   I(X_\oo ; C) \right] - \Delta I(X_\NN; C)
\label{eq:6.7}
}

By conservation of energy, the work done on the system during $[t_i, t_f]$ is the change in its internal energy 
minus the heat flow to all the reservoirs, which is given by sum of the (temperature normalized) entropy flows
to the reservoirs. (Equivalently, this is the negative of the work extracted from the system.)
Since the Hamiltonian is uniform at $t_i, t_f$, the change in internal energy is zero.
For simplicity assume all reservoirs have the same temperature, $T$, and choose units so that
$k_B T = 1$. Then that sum of entropy flows is the total change in the entropy of the system minus the EP.

Combining this with \cref{eq:height_2},  it is shown in App.\,J that the amount of work that can be extracted from the system
under feedback control if one takes into account the unit structure is (perhaps loosely) upper-bounded by
\eq{
\Delta \left[\SSS_{\oo \in {\NN^*}}  S(X_\oo | C) \right] 
\label{eq:6.10}
}
In contrast, the conventional analysis in the literature,
in which one does not account for the unit structure of the system, results in an upper bound
of $\Delta S(X_\NN | C)$~\cite{sagawa2009minimal,parrondo2015thermodynamics, kolchinsky_wolpert_2020_thermo_under_constraints}.
The difference between these two terms is how much the unit structure restricts the amount of work
we can extract from a system by observing its state.






\section{Examples of the strengthened second law}
\label{sec:examples}

In this section I work through some elementary examples illustrating \cref{eq:height_2}.
All unit structures in these examples are implicitly assumed to have height less than $3$.


\subsection{Example 1}

Consider any process where every coordinate that is in the intersection of two or more distinct units stays constant throughout
the process. 
In such a process
\eq{
-\Delta \II^{\NN^*} = \left[ \sum_{\oo \in \Gamma_{\NN^*}^R}S(p_\oo(t_i)) - S(p_\oo(t_f))\right] - \left[S(p(t_i)) - S(p(t_f))\right] 
\label{eq:23}
}
where the sum runs only over the root nodes.
%
Moreover, since any unit that never changes its state generates no EP, in this kind of process
\eq{
{\SSS}_\oo \sigma^\oo &= \sum_\oo \sigma^\oo
\label{eq:6.2}
}
by the definition of in-ex sum.
The lowest each $\sigma^\oo$ can be 
is zero (which occurs when each unit $\oo$ evolves semi-statically slowly). Therefore we can combine \cref{eq:6.2} with \cref{{eq:12,eq:23}} to establish that
the lower bound on EP is
 \eq{ 
\BBB_{\NN^*} = \left[ \sum_{\oo}S(p_\oo(t_i)) - S(p_\oo(t_f))\right] - \left[S(p(t_i)) - S(p(t_f))\right] 
}
exactly, i.e., \cref{eq:23} is a strict lower bound on the EP. This lower bound holds no matter what $p_{t_i}(x)$ and $p(x({t_f}) \,|\, x(t_i))$ are, so long as $p(x({t_f}) \,|\, x(t_i))$  is consistent with the unit structure. 

As an illustration of this result, suppose that no two units intersect one another, and that every unit contains just a single coordinate. 
Then the lower bound on EP is  
\eq{
\BBB_{\NN^*} =  \left[\sum_{i} S(p_i(t_i)) - S(p_i(t_f))\right] - \left[S(p(t_i)) - S(p(t_f))\right]
\label{eq:24}
}
which is the drop among the coordinates in their multi-information, sometimes called ``total correlation''.  (This lower bound
on the EP was previously derived in~\cite{wolpert_thermo_comp_review_2019,kolchinsky_wolpert_2020_thermo_under_constraints},
in the special case that each coordinate is a physically separate subsystem.)
By repeated application of the data-processing inequality, it is easy to confirm that this lower bound on the EP is non-negative.

Note though that \cref{eq:23} holds for \textit{any} process with a height $2$ unit structure,
so long as the ending entropies of (the joint coordinates in the units corresponding to) the leaf nodes equal the 
associated starting entropies. In particular, this is true even if the coordinates in the leaf nodes \textit{do} change state during the process.
Since the dependency graph has height $2$, \cref{eq:height_2}
tells us that the expression in \cref{eq:23} is a lower
bound on the EP of such a process. Furthermore, the same argument using the data-processing inequality establishes that that lower bound
is non-negative. However, in general, if the coordinates in the leaf nodes change their states during
the process, that lower bound may not be tight.

\subsection{Example 2}

Suppose that a system comprises three physically separated subsystems, $\{1, 2, 3\}$,
each with two possible states, $0$ and $1$. Suppose as well that the dynamics
can be represented with the height-$2$ unit structure ${\AAA^*} = \{\{1,2\}, \{2\}, \{2, 3\}\}$. So
subsystem $2$ evolves independently, while the dynamics of both
subsystems $1$ and $3$ depend on the state of subsystem $2$.

Suppose as well that 
initially, $x_1 = x_3$ with uniform probability over their two possible joint states, and
that $x_2$ is independent of both $x_1$ and $x_3$, also with uniform probability over its states:
\eq{
p_{t_i}(x) &= \dfrac{1}{4} \sum_{k =0}^1 \delta(x_1(t_i), k)  \delta(x_3(t_i), k) \sum_{m =0}^1 \delta(x_2(t_i), m)
} 
Therefore $S(p(t_i)) = 2\ln 2$, and so
\eq{
\II^{{\AAA^*}}(p(t_i)) &= \left[2 \ln 2 + 2 \ln 2 - \ln 2\right] - 2\ln 2 \\
	&= \ln 2
\label{eq:29}
}

Assume that $x_2$ eventually loses all information about its initial state. So 
\eq{
p \left(x_2(t_f)\,|\, x(t_i) \right) = p \left(x_2(t_f)\,|\, x_2(t_i)\right) = p \left(x_2(t_f)\right)
}
In addition, as required by the unit structure, have $x_1$ and $x_3$ evolve independently of one another, conditioned on the state $x_2$, 
and presume that they both eventually lose all information about their own initial states and the initial state of $x_2$. 
So for example,
\eq{
 p \left(x_1(t_f) \,|\, x_2(t_f), x_1(t_i), x_2(t_i) \right) =  p \left(x_1(t_f) \,|\, x_2(t_f)\right)
}
Combining,

\eq{
p\left(x(t_f) \,|\, x(t_i) \right) &= p \left(x_1(t_f), x_3(t_f) \,|\, x_2(t_f), x(t_i) \right) \, p \left(x_2(t_f)\,|\, x(t_i) \right) \\
	&= p \left(x_1(t_f) \,|\, x_2(t_f), x_1(t_i), x_2(t_i) \right) \, p \left(x_3(t_f) \,|\, x_2(t_f), x_2(t_i), x_3(t_i) \right) \, p \left(x_2(t_f) \right) \nonumber \\
	&= p \left(x_1(t_f) \,|\, x_2(t_f)\right) \, p \left(x_3(t_f) \,|\, x_2(t_f)\right) \, p \left(x_2(t_f) \right) 
\label{eq:30}
}
Therefore 
$S(X(t_f)) = 
S(X_1(t_f) \,|\, X_2(t_f) ) + S(X_3(t_f) \,|\, X_2(t_f)) + S(X_2(t_f) ) 
$,
and so
\eq{
\II^{{\AAA^*}}(p(t_f)) &= - S(X(t_f)) \nonumber \\
&\!\!\!\!\!\!\!\!+   \left[ \left(S(X_1(t_f) \,|\, X_2(t_f)) + S(X_2(t_f)) \right) + \left(S(X_3(t_f) \,|\, X_2(t_f)) + S(X_2(t_f)) \right) - S(X_2(t_f))\right] 
				\nonumber \\
	&= 0
\label{eq:37}
}


Combining \cref{eq:29,eq:37} establishes that the EP is lower-bounded by $\ln 2$. Note that we can derive this
lower bound on the EP even though both subsystems $1$ and $3$ are continually observing subsystem $2$ during the process,
even if subsystem $2$'s state is changing as they observe it. 
In addition, this lower bound holds no matter what the ending distribution $p_{t_f}(x)$ is,
so long it can be written as in \cref{eq:30}.
(So in particular, as discussed in the introduction, it applies to a simple extension of the 
cell-sensing scenario analyzed in~\cite{hartich_sensory_2016,bo2015thermodynamic}.)

\subsection{Example 3}
\label{ex:5}

\begin{figure}[tbp]
\includegraphics[width=75mm]{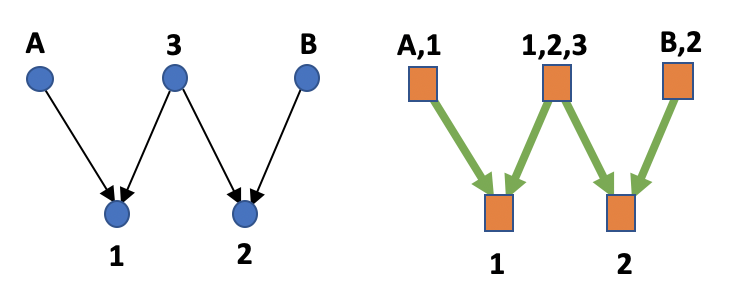}
\caption{The random walker scenario described in the introduction and investigated in Ex.\,$3$. a) In the left
panel the five coordinates
are indicated by circles, with the associated rate matrix dependencies indicated by arrows, using the
same convention as in \cref{fig:1}. b) The right panel shows a height-$2$ dependency graph
for this rate matrix. Each square is a different unit, with the associated coordinates explicitly written. Note that in dependency graphs arrows
indicate the partial order of subset inclusion. In this example, the number of units is the same as the number of coordinates, 
but that need not be the case in general.
}
\label{fig:4}
\end{figure}

Return to the example of a random walker presented in the introduction, with an associated
height $2$ unit structure illustrated in  \cref{fig:4}. Plugging into \cref{eq:in_ex_info} and using obvious shorthand,
\eq{
-\Delta \II^{\NN^*} &= \Delta \bigg[S(\NN) - S(1) - S(2) - S({1,2,3}) - S({A,1}) - S({B,2}) \nonumber \\
	& \qquad\qquad\qquad + \left(S(1) + S(1) + S(1) + S(2) + S(2) + S(2)\right)\bigg] \\
	&= \Delta \bigg[S(A, B \,|\, 1, 2, 3) + S(1) + S(2) - S({A \,|\,1}) - S({B \,|\, 2})\bigg]
}

Suppose that at $t_i$ the full system has a single specific state with probability $1$. So $S(t_i) = 0$.
Suppose as well that the position in the lattice is uniformly random at $t_f$. (For example, 
this will occur at large enough $t_f$ if the lattice has periodic boundary conditions and both 
$X_1$ and $X_2$ evolve by randomly choosing one of their two neighbors.) 
This means that knowing the values of $x_2, x_3$ at $t_f$ tells us nothing about the most recent
values of $x_1$ not already given by the value of $x_1$ at $t_f$, and so in particular tells us
nothing about the most likely value of $x_A$ then. The same is true concerning the value of $x_B$ at $t_f$.
This all means that $S_{t_f}(A, B \,|\, 1, 2, 3) = S_{t_f}(A \,|\, 1) + S_{t_f}(B \,|\, 2)$. 

Combining gives $-\Delta \II^{\NN^*} \;=\; 2 \ln L$.
So \cref{eq:height_2} provides a strictly positive lower bound on the global EP. Note that this lower bound applies
no matter what the dynamics of the process; it can be quasi-statically slow, it can involve Hamiltonian quenches,
but so long as the unit structure does not change during the process, the EP is lower-bounded by $2 \ln L$. 
Furthermore, so long as the Hamiltonian is uniform at both $t_i$ and $t_f$, the total work extracted in
the process is the gain in entropy of the full system minus the global EP. Combining establishes that
the total work extracted is upper-bounded by 
\eq{
S_{t_f}(A \,|\, 1) + S_{t_f}(B \,|\, 2) + 2 \ln N
\label{eq:5.18}
}

Note that increasing $N$ while keeping $LN$ constant means that the precise value $(x_1, x_2)$ tells us less
about the precise lattice position.  \cref{eq:5.18} tells us that
increasing the significance of $x_3$ this way increases the upper bound on the total amount of work that can be extracted.

\section{Discussion}

In this paper I consider the thermodynamics of multi-dimensional systems
evolving according to a continuous-time Markov chain. My main result is a strengthened version of the conventional second law,
which applies whenever we have an \textit{a priori} set of ``dependency constraints'' that
for each coordinate $i$ specify which other coordinates can directly affect the dynamics of $i$, via the rate matrix.
The result holds for any coordinate system  --- the coordinates
can be conventional phase space coordinates, they can be states of a set of
separate interacting subsystems of an overall system, they can be 
positions in a sequence of more refined coarse-grainings of the state of the system, they can involve amounts of
various chemicals in the system, etc. 

To derive my result I first translate the dependency constraints
into a  ``unit structure''. This gives a sigma algebra that groups the coordinates into overlapping sets, in a way that
respects the dependency constraints. In general, any set of dependency constraints can be translated into
more than one unit structure. In turn, any unit structure specifies an information-theoretic functional of 
distributions over the states of the system, called the ``in-ex information''. To illustrate this, suppose the dependency
constraints specify that each
coordinate evolves autonomously, independent of the others. (As an example, this would be the case
for the spatial coordinates of a particle freely evolving under over-damped Langevin dynamics in a uniform medium with no
external forces.) We could then choose a unit structure that assigns each
coordinate to its own unique unit. In this case the in-ex information reduces to the
total correlation (sometimes called ``multi-information'') of the system's distribution,
with each coordinate viewed as a separate random variable. 

The strengthened version of the second law derived in this paper says that the entropy production (EP)
of the system is lower-bounded by the difference between the beginning
and ending values of the system's in-ex information. This lower bound 
is independent of all features of the dynamics other than the the beginning distribution, the ending distribution, and the
dependency constraints restricting how the dynamics could have caused the initial distribution to evolve
into the ending distribution. Accordingly, we can use this strengthened second law to upper-bound the amount of work that
can be extracted from a system as it evolves from one specified distribution to
another~\cite{parrondo2015thermodynamics,hasegawa2010generalization},
in a way that accounts for dependency constraints governing the system's dynamics. Similarly, this strengthened second law can be
used to refine recent results in thermodynamics of feedback control~\cite{sagawa2010generalized},
to account for dependency constraints in the system being controlled.

In contrast to
other similar recently derived lower bounds on EP~\cite{wolpert2020fluctuation,wolpert2020minimal},
the one derived here does not require that the dynamics of the system be a multi-partite process. Nor does
it require that local detailed balance holds. These two features mean the lower bound applies
to any system undergoing continuous-time Markovian dynamics, even if the system has no
natural thermodynamic interpretation. As a result, we can apply these results to everything from (Markov models of)
evolving opinion networks to replicator dynamics of a population of evolving organisms.

A recent paper~\cite{kolchinsky_wolpert_2020_thermo_under_constraints} used an
information-geometric analysis to also derive bounds on
minimal entropy production (EP) that arise due to constraints on the rate matrix of a system's dynamics.
%
%
%
%
To use the analysis in~\cite{kolchinsky_wolpert_2020_thermo_under_constraints} one needs to first 
find an operator $\phi$ over the set of all joint distributions which
both obeys the Pythagorean theorem of information theory and which commutes with the time-evolution
operators defined by the set $\Lambda$ of allowed rate matrices.  In general there are many such $\phi$, but different ones
will result in different bounds on EP. 

The analog of $\Lambda$ in this
paper is the set of dependency constraints.The analog to finding one (or more) $\phi$'s for the approach
in this paper is choosing a coordinate system and associated unit structure that represents the dependency constraints and is
rich enough for the lower bound on EP to be strictly positive. Similarly to the case with the
approach in~\cite{kolchinsky_wolpert_2020_thermo_under_constraints}, where different $\phi$ all consistent with the constraints on the set of allowed rate matrices will result in different bounds on EP, in general different unit structures
all consistent with the constraints on the set of allowed rate matrices will result in different bounds on EP. 


\cite{kolchinsky_wolpert_2020_thermo_under_constraints} provides many examples of how 
constraints on the allowed rate matrices can be used to derive nonzero lower bounds on EP,
including collective flashing ratchets,
Szilard boxes where the particle is subject to a gravitational force in addition to driving by a piston,
Szilard boxes where there are constraints on the way the piston can be used, evolving Ising spin systems, etc.
Many of these examples can be formulated as systems evolving under dependency constraints (e.g., most
of the examples involving in~\cite{kolchinsky_wolpert_2020_thermo_under_constraints} ``modularity constraints'' 
can be directly formulated this way). 
Future work involves comparing the EP bounds in this paper to the ones
in~\cite{kolchinsky_wolpert_2020_thermo_under_constraints}, and more generally trying to 
synthesize the two approaches.

$ $

\ack{ I would like to thank G{\"u}lce Kardes, Artemy Kolchinsky and
especially Farita Tasnim for feedback on drafts of this paper. 
This work was supported by the Santa Fe Institute, 
Grant No. CHE-1648973 from the US National Science Foundation and Grant No. FQXi-RFP-IPW-1912 from the FQXi foundation.
The opinions expressed in this paper are those of the author and do not necessarily 
reflect the view of the National Science Foundation.}


\enlargethispage{20pt}

\bibliographystyle{amsplain}
\bibliography{../../../../BIB/thermo_refs.main.3.BIB.DIR,new.motivation.strengthened.second.law.rsta.version.main.POST.RESPONSE.TO.REFEREES.v2.bib}

\newpage
\break

\section*{\huge{Supplementary material for  \textit{Strengthened second law for multi-dimensional systems
coupled to multiple thermodynamic reservoirs}}}

$ $

$ $

\noindent \Large{David H. Wolpert$^1$}

$ $

\noindent \normalsize{$^{1}$Santa Fe Institute, New Mexico, USA}

\noindent {Complexity Science Hub, Vienna}

\noindent {Arizona State University, Tempe, Arizona}

\noindent {International Center for Theoretical Physics, Italy}

\noindent \texttt{http://davidwolpert.weebly.com}

$ $

\appendix

\renewcommand{\theequation}{\thesection\arabic{equation}}

\section{Proof that the intersections of two units is a unit}
\label{app:unit_intersections}

Without loss of generality, suppose that there are exactly three coordinates, with their values written as
$x, y, z$, respectively, and hypothesize that the joint state $(x, y)$ evolves as a unit, as does $(y, z)$. We need to prove that $y$ also
evolves on its own as a unit.

Let $K^{x',y',z'}_{x,y,z}$ be the overall rate matrix of the system at (implicit) time $t$. By our
two hypotheses, there are two rate matrices $F, G$, such that
\eq{
\sum_{z,z'} \sum_{x',y'} K^{x',y',z'}_{x,y,z} p_{x',y',z'} &= \sum_{x',y'} F^{x',y'}_{x,y}  p_{x',y'} \\
\sum_{x,x'} \sum_{z',y'} K^{x',y',z'}_{x,y,z} p_{x',y',z'} &= \sum_{z',y'} G^{z',y'}_{z,y}  p_{z',y'} 
}
Since these equations must hold for all $p$, by considering delta function $p$'s we see that
\eq{
\label{eq:A1}
\sum_z K^{x',y',z'}_{x,y,z} &= F^{x',y'}_{x,y} \qquad \forall x', y', z', x, y \\
\sum_x K^{x',y',z'}_{x,y,z} &= G^{z',y'}_{z,y} \qquad \forall x', y', z', z, y 
\label{eq:aa2}
}

Next, define
\eq{
W^{y',x'}_y &:= \sum_x  F^{x',y'}_{x,y}
\label{eq:aa3}
}
Summing both sides of \cref{eq:aa1} over $x$ and then plugging in \cref{eq:aa2},
\eq{
\sum_x  F^{x',y'}_{x,y} &= \sum_{x,z} K^{x',y',z'}_{x,y,z}   \\
	&= \sum_z  G^{z',y'}_{z,y}
}
Since the RHS is independent of $x'$, so must the LHS be. So $W^{y',x'}_y$
must be independent of $x'$. Moreover, combining \cref{eq:aa3,eq:aa2} shows
that $\sum_y W^{y',x'}_y = 0$ for all $y'$. Therefore $U^{y'}_y := W^{y',x'}_y$
is a properly normalized rate matrix. 

To complete the proof, expand
\eq{
\dfrac{d}{dt} p_y &= \sum_{x,z} \dfrac{d}{dt} p_{x,y,z} \\
  &=	 \sum_{x,z} \sum_{x',y',z'} K^{x',y',z'}_{x,y,z} p_{x',y',z'} \\
  &= \sum_{x',y'} W^{y',x'}_y p_{x',y'} \\
	&= \sum_{x',y'} U^{y'}_y p_{x',y'} \\
	&= \sum_{y'} U^{y'}_y p_{y'} 
}

\section{Illustration of 
\cref{eq:consistency_rate_matrices}}
\label{app:illustration}

Suppose that each coordinate of $x$ has the same dimension, $m$. 
Choose any subset of the coordinates, $\alpha$,
coordinate $i \in \alpha$ and any stochastic matrix $W^{x'_i}_{x_i}$. Then 
\eq{
K^{x'}_x :=  W^{x'_i}_{x_i} {\prod}_{j \ne i}\delta(x_i, x_j)
}
is a properly normalized rate matrix over all of $X$, since for all $x'$,
$\sum_x K^{x'}_x = 
\sum_{x_i} W^{x'_i}_{x_i} = 0$. 

In addition, choose $\oo = \NN$, and choose
\eq{
K^{x'_\alpha}_{x_\alpha}(\alpha) :=   W^{x'_i}_{x_i} {\prod}_{j \in \alpha : j \ne i}\delta(x_i, x_j) 
}
Then $K^{x'}_x$ obeys \cref{eq:consistency_rate_matrices}
in the main text, since for all $x_\alpha, x_{-\alpha}, x'_\alpha, x'_{-\alpha}$,
\eq{
 \sum_{x_{-\alpha}} K^{x'_\alpha, x'_{-\alpha}}_{x_\alpha, x_{-\alpha}} &=  \sum_{x_{-\alpha}} W^{x'_i}_{x_i} {\prod}_{j \in \alpha : j \ne i}\delta(x_i, x_j)
{\prod}_{j \not\in \alpha}\delta(x_i, x_j)
 \\
	&=  K^{x'_\alpha}_{x_\alpha}(\alpha)
}
So $\alpha$ is a unit.

\section{Proof of 
\cref{eq:3.6}}
\label{app:can_ignore_what_not_in_leaderset}

Expand
\eq{
\langle \dot{Q}(t) \rangle &= \sum_{v, x', x} L^{x'}_x(v; t) p_{x'}(t) \ln \dfrac{L^{x'}_{x}(v; t)}{L^{x}_{x'}(v; t)} \\
	&= \sum_{v} \sum_{x'_{\LL(v)},\,x'_{-\LL(v)}} \sum_{x_{\PP(v)},\,x_{-\PP(v)}} 
	L^{x'_{\LL(v)}}_{x_{\PP(v)}}(v; t)  \delta^{x'_{-\PP(v)}}_{x_{-\PP(v)}} p_{x'_{\LL(v)},x'_{-\LL(v)}}(t) 
			\ln \dfrac{L^{x'_{\LL(v)}}_{x_{\PP(v)}}(v; t) \delta^{x'_{-\PP(v)}}_{x_{-\PP(v)}}} {L^{x_{\LL(v)}}_{x'_{\PP(v)}}(v; t) \delta^{x_{-\PP(v)}}_{x'_{-\PP(v)}}} \\
	&= \sum_{v} \sum_{x'_{\LL(v)},\,x'_{-\LL(v)}} \sum_{x_{\PP(v)},\,x_{\LL(v) \setminus \PP(v)}} 
			 L^{x'_{\LL(v)}}_{x_{\PP(v)},x'_{\LL(v) \setminus \PP(v)}}(v; t) \nonumber \\
	&\qquad  \times \;  \delta^{x'_{-\LL(v)}}_{x_{-\LL(v)}} p_{x'_{\LL(v)},x'_{-\LL(v)}}(t) 
			\ln \dfrac{L^{x'_{\LL(v)}}_{x_{\PP(v)},x'_{\LL(v) \setminus \PP(v)}}(v; t) \delta^{x'_{-\LL(v)}}_{x_{-\LL(v)}}}   {L^{x_{\LL(v)}}_{x'_{\PP(v)},x_{\LL(v) \setminus \PP(v)}}(v; t) \delta^{x_{-\LL(v)}}_{x'_{-\LL(v)}}} \\
	&= \sum_{v} \sum_{x'_{\LL(v)},\,x'_{-\LL(v)}} \sum_{x_{\LL(v)} }
	L^{x'_{\LL(v)}}_{x_{\PP(v)},x'_{\LL(v) \setminus \PP(v)}}(v; t)     p_{x'_{\LL(v)},x'_{-\LL(v)}}(t) 
			\ln \dfrac{L^{x'_{\LL(v)}}_{x_{\PP(v)},x'_{\LL(v) \setminus \PP(v)}}(v; t)}   {L^{x_{\LL(v)}}_{x'_{\PP(v)},x_{\LL(v) \setminus \PP(v)}}(v; t) }  \\
	&= \sum_{v} \sum_{x'_{\LL(v)},} \sum_{x_{\LL(v)}} 
	L^{x'_{\LL(v)}}_{x_{\PP(v)},x'_{\LL(v) \setminus \PP(v)}}(v; t)   p_{x'_{\LL(v)}}(t) 
			\ln \dfrac{L^{x'_{\LL(v)}}_{x_{\PP(v)},x'_{\LL(v) \setminus \PP(v)}}(v; t)}   {L^{x_{\LL(v)}}_{x'_{\PP(v)},x_{\LL(v) \setminus \PP(v)}}(v; t) }  \\
	&= \sum_{v} \sum_{x'_{\LL(v)}} \sum_{x_{\LL(v)}} 
	L^{x'_{\LL(v)}}_{x_{\PP(v)}}(v; t)  p_{x'_{\LL(v)}}(t) 
			\ln \dfrac{L^{x'_{\LL(v)}}_{x_{\PP(v)}}(v; t) } {L^{x_{\LL(v)}}_{x'_{\PP(v)}}(v; t)} 
%
%
}
where for convenience I have switched back and forth between the shorthand of~\cref{eq:2.8}.

\section{Proof of 
\cref{eq:13}}
\label{app:union_increases_EP}

First, note that by \cref{eq:consistency_rate_matrices},
for all
conditional distributions $p(x'_{\oo \setminus \alpha} \,|\, x'_\alpha)$, for all  $x_\alpha, x'_\alpha$, 
\eq{
 K^{x'_\alpha}_{x_\alpha}(\alpha; t) &=  \sum_{x_{\oo \setminus \alpha}, x'_{\oo \setminus \alpha}} 
			K^{x'_\alpha, x'_{\oo \setminus \alpha}}_{x_\alpha, x_{\oo \setminus \alpha}}(\oo; t) p(x'_{\oo \setminus \alpha} \,|\, x'_\alpha) \\
	&=  \sum_{v \in\nu(\oo)} \sum_{x_{\oo \setminus \alpha}, x'_{\oo \setminus \alpha}} 
			L^{x'_\alpha, x'_{\oo \setminus \alpha}}_{x_\alpha, x_{\oo \setminus \alpha}}(v; t) p(x'_{\oo \setminus \alpha} \,|\, x'_\alpha)
\label{eq:consistency_rate_matrices2}
}
where the second equality follows from \cref{app:proof_prop:1b}.
Next, use 
\cref{eq:5} 
to expand
\eq{
\dSot &= \sum_{v \in \nu(\oo)} \sum_{x_\alpha, x'_\alpha} p_{x'_\alpha}(t) 
		 \sum_{x_{\oo\setminus \alpha}, x'_{\oo\setminus \alpha}}  L^{x'_\alpha,x'_{\oo \setminus \alpha}}_{x_\alpha,x_{\oo \setminus \alpha}} (v;t)
    p_t(x'_{\oo \setminus \alpha} \,|\, x'_\alpha)  \nonumber \\
&\qquad	\qquad \times	\left( \ln \left[\dfrac{L^{x'_\alpha,x'_{\oo \setminus \alpha}}_{x_\alpha,x_{\oo \setminus \alpha}} (v;t)  
								p_t(x'_{\oo \setminus \alpha} \,|\, x'_\alpha)}   
					 {L^{x_\alpha,x_{\oo \setminus \alpha}}_{x'_\alpha,x'_{\oo \setminus \alpha}} (v;t)  p_t(x_{\oo \setminus \alpha} \,|\, x_\alpha)} \right] 
						+ \ln \left[\dfrac{p_{x'_\alpha}}{p_{x_\alpha}} \right] 	\right)
\label{eq:aa1}
}

In addition, using the log sum inequality~\cite{cover_elements_2012} shows that for each $v, x_\alpha, x'_\alpha$, the associated value of
the first term in the inner sum on the RHS of \cref{eq:aa1} is bounded by
\eq{
&\!\!\!\!\!\!\!\!\! \sum_{x_{\oo\setminus \alpha}, x'_{\oo\setminus \alpha}}  L^{x'_\alpha,x'_{\oo \setminus \alpha}}_{x_\alpha,x_{\oo \setminus \alpha}} (v;t)
    p_t(x'_{\oo \setminus \alpha} \,|\, x'_\alpha) 
	\ln \left[\dfrac{L^{x'_\alpha,x'_{\oo \setminus \alpha}}_{x_\alpha,x_{\oo \setminus \alpha}} (v;t)  p_t(x'_{\oo \setminus \alpha} \,|\, x'_\alpha)}   
					 {L^{x_\alpha,x_{\oo \setminus \alpha}}_{x'_\alpha,x'_{\oo \setminus \alpha}} (v;t)  p_t(x_{\oo \setminus \alpha} \,|\, x_\alpha)} \right]   \nonumber \\
		& \qquad \ge 
\left[\sum_{x_{\oo\setminus \alpha}, x'_{\oo\setminus \alpha}}  L^{x'_\alpha,x'_{\oo \setminus \alpha}}_{x_\alpha,x_{\oo \setminus \alpha}} (v;t)
    p_t(x'_{\oo \setminus \alpha} \,|\, x'_\alpha) \right]
 	\ln \dfrac{\left[ \sum_{x_{\oo\setminus \alpha}, x'_{\oo\setminus \alpha}} L^{x'_\alpha,x'_{\oo \setminus \alpha}}_{x_\alpha,x_{\oo \setminus \alpha}} (v;t)  p_t(x'_{\oo \setminus \alpha} \,|\, x'_\alpha) \right]}   
					 {\left[ \sum_{x_{\oo\setminus \alpha}, x'_{\oo\setminus \alpha}}  L^{x_\alpha,x_{\oo \setminus \alpha}}_{x'_\alpha,x'_{\oo \setminus \alpha}} (v;t)  p_t(x_{\oo \setminus \alpha} \,|\, x_\alpha) \right]}  \nonumber \\
		& \qquad =  L^{x'_\alpha}_{x_\alpha}(v; t) \ln \left[\dfrac{ L^{x'_\alpha}_{x_\alpha}(v; t)}{ L^{x_\alpha}_{x'_\alpha}(v; t)}\right]
}
where the second line uses \cref{eq:consistency_rate_matrices2}.

Combining and using \cref{eq:consistency_rate_matrices2} again establishes that 
\eq{
\dSot & \ge \sum_{v \in \nu(\alpha)} \sum_{x_\alpha, x'_\alpha} L^{x'_\alpha}_{x_\alpha}(v; t) p_{x_\alpha, x'_\alpha}(t) \ln \left[\dfrac{ L^{x'_\alpha}_{x_\alpha}(v; t)}{ L^{x_\alpha}_{x'_\alpha}(v; t)}\right]
}
Again plugging into  Eq.\,3.13,
this time for the EP  rate of unit $\alpha$, completes the proof.

\section{Proof of 
\cref{eq:11aa}}
\label{app:mobius}

We are given some composite system with unit structure $\NN^*$.
Fix the time $t$ and make it implicit for the rest of this appendix. Define $\mathcal{T}$ as the set of all 
state transitions in $X$, involving an arbitrary number
of the coordinates in $\NN$. (So there are $|X| \left(|X| - 1\right)$ elements of $\mathcal{T}$,
where $|X| = \prod_{i \in \NN} |X_i|$ is the number of joint states of the multi-dimensional system.)

For any unit $\oo \in \NN^*$
define $\tau_\omega$ as the set of all state transitions $x' \rightarrow x \ne x'$ such that both $K^{x'}_x \ne 0$ and
$x_{-\oo} = x'_{-\oo}$. (Intuitively,  $\tau_\omega$ is the set of all state transitions that do not modify any of the
coordinates outside of $\oo$ and that are allowed under the CTMC.)
Note that $\tau_\oo \subseteq \tau_{\oo'}$ if $\oo \subseteq \oo'$. So
${\mathcal{T}}_{\NN^*} := \{\tau_\omega : \oo \in \NN^*\} \cup \{\mathcal{T}\}$
is a locally finite poset, 
ordered by the set inclusion relation.


Due to our assumption that the unit structure is flush, no state transition $x' \rightarrow x$ can occur 
that has the property of simultaneously changing the state of all coordinates 
in an arbitrary set $\alpha$ unless $\alpha$ is a subset of a unit $\oo \in \NN^*$
(i.e., there is no pair $(x', x)$ that has that property where both $p_{x'} \ne 0$ and $K^{x'}_x \ne 0$). This means that
every element of $\mathcal{T}$ allowed by the CTMC is contained in at least one element of ${\mathcal{T}}_{\NN^*}$. So ${\mathcal{T}}_{\NN^*}$ is
a cover of the set of all state transitions allowed by the CTMC. 

Next, due to our assumption that there are no two equivalent units in the unit structure,
the set of all transitions in a set $\tau_\oo$ uniquely specifies $\oo$, i.e., there is a bijection
between $\Omega := \NN^* \cup \NN$ and $\mathcal{T}_{\NN^*}$. Due to this, without any ambiguity we can define
a function $g : \mathcal{T}_{\NN^*} \rightarrow \R$ by setting
\eq{
g(\tau_\oo)      &:= \sum_{v \in \nu(\oo)} \sum_{(x',x) \in \tau_\oo} L^{x'_\oo}_{x_\oo}(v) p_{x'_\oo} \ln \left[\dfrac{L^{x'_\oo}_{x_\oo}(v)}
				{L^{x_\oo}_{x'_\oo}(v)}\right]   \\
    &=  \sum_{x'_\oo,x_\oo} L^{x'_\oo}_{x_\oo}(v) p_{x'_\oo} \ln \left[\dfrac{L^{x'_\oo}_{x_\oo}(v)}
				{L^{x_\oo}_{x'_\oo}(v)}\right]   \\
&= \langle \dot{Q}^\oo \rangle 
}
for all $\oo \in \NN^*$,
and similarly setting $g(\mathcal{T}) = \langle \dot{Q}^\NN \rangle$. (Note that if $\NN \in \NN^*$, then this second definition
is redundant.)


In addition, unit structures are closed under intersection. As described in the text, 
this means that $\NN^*$ generates a sigma algebra over coordinates, and we can use
the function $\{\langle \dot{Q}^\oo \rangle : \oo \in \NN^*\}$ to generate a signed measure over that sigma algebra.
In the same way, $\mathcal{T}_{\NN^*}$ generates a sigma algebra over state transitions, and the
function $g(\tau_\oo)$ generates a signed measure of that sigma algebra.
We can apply all the steps in the usual derivation of the inclusion-exclusion principle for signed measures
from Rota's theorem~\cite{stanley2011enumerative} to that signed measure generated by $g(.)$. This allow us to write
\eq{
g(\mathcal{T}) &= \SSS_{\oo \in {\NN^*}} g(\tau_\omega)
}

Plugging in the definition of $g(.)$
completes the proof.

\section{Discussion of Eq.\,3.21
for the case of parallel bit erasure}
\label{app:bit_erasure_example}

Suppose our system comprises two physically separated subsystems 
where that state space of both subsystems are a bit. Suppose as well that  subsystem $1$ gets erased in the process, 
and the initial distribution is
uniform random between the joint state where both bits equal $0$, and the joint state where both bits equal $1$, with no other possibilities. 
If the dynamics is a single
unit, so that the rate matrix of subsystem $1$ can depend on the (unchanging) state of subsystem $2$, EP can equal $0$.
If instead there are two separate units, so that the rate matrix of subsystem $2$ cannot depend on the state of subsystem $2$,
minimal EP is instead $\ln 2$. 

To understand this, suppose that there is a (time-varying) Hamiltonian and that the full system's  (time-varying) rate matrix  always obeys LDB for that
Hamiltonian. So to have the global EP during the full process equal zero, there would have to be a trajectory of such Hamiltonians such
that at all times the system was always at equilibrium for the associated Hamiltonian. Furthermore, because we're assuming the system
is governed by a CTMC,  the distribution $p_{x_1,x_2}(t)$ can never change discontinuously during times $t \in (0, 1)$ within the
process, and therefore neither can the Hamiltonian. 

Given this, suppose that there is no
continuous trajectory of such distributions that is always a product distribution and has the desired initial and final forms, $p_{x_1,x_2}(t_i), p_{x_1,x_2}(t_f)$.
(For example, this is the case if the two subsystems are statistically coupled at $t = 0$.)
Then by the always-at-equilibrium condition for zero EP, there must be a time $t \in (0, 1)$ at which the Hamiltonian cannot be written as
a sum of a function of $x_1$ plus a function of $x_2$, but instead must nonlinear couple them. By our assumption of LDB, this would then mean that at that time $t$,
the full rate matrix of the joint system must couple those two subsystems, which in turn means that the rate matrix
of one of the two subsystems must depend on the state of the other subsystem \footnote{Note that how much EP in a process is
generated is independent of whether the system obeys LDB, or even whether there is a Hamiltonian. On the other hand,
if there \textit{is} a Hamiltonian that obeys LDB, we know that zero EP would be generated iff the system were always
at equilibrium.}

To apply this to our two-subsystem example, simply note that the beginning distribution is not a
product distribution. Therefore the condition for zero EP is violated if neither of the rate matrices of the two subsystems depends
on the other subsystem's state.

\section{Proofs related to Eq.\,4.1}

\label{app:second_major_result}

To begin, I need to
define $V(\NN^*)$, a set of distributions over the Boolean hypercube, ${\B^{|\NN^*|}}$.
After that, I define a ``centering distribution'' to be any convex combination of the elements of $V(\NN^*)$
which equals $(1 / |\NN^*|, 1 / |\NN^*|, \ldots)$, the uniform distribution over the Boolean hypercube.
My first main result, presented in \cref{prop:1}, is a function taking each such centering distribution to a different lower bound on global EP.


The set $V(\NN^*)$ is the union of two sets of distributions over $\B^{|\NN^*|}$,
which I define in succession:

$ $

\noindent I)
For any $\oo \in \NN^*$, write $\delta^\oo$ for the distribution over $\B^{|\NN^*|}$ which is all $0$'s except for
a $1$ in its $\oo$ component. 
Using this notation, define $V^1(\NN^*)$ as the set of all distributions $\alpha$ over ${\B^{|\NN^*|}}$ which obey at least one of the following three conditions:
\begin{enumerate}
\item $\alpha = \delta^\oo$ for some $\oo \in \NN^*$ whose height $\le 2$;
\item $\alpha = \dfrac{\sum_{\oo' \in \family(\oo)} \delta^{\oo'}}{\sum_{\oo' \in \family(\oo)} 1}$ for some $\oo \in \NN^*$;
\item $\alpha = \dfrac{\sum_{\oo' \in \family(\oo) \setminus \family(v)} \delta^{\oo'}}{\sum_{\oo' \in \family(\oo) \setminus \family(v)} 1}$ for a $\oo \in \NN^*, v \in \desc(\oo)$;
\end{enumerate}
I will refer to any distributions that obey these conditions as \textbf{type-$1$}, \textbf{type-$2$}, and \textbf{type-$3$} distributions, respectively.

\begin{figure}[tbp]
\includegraphics[width=75mm]{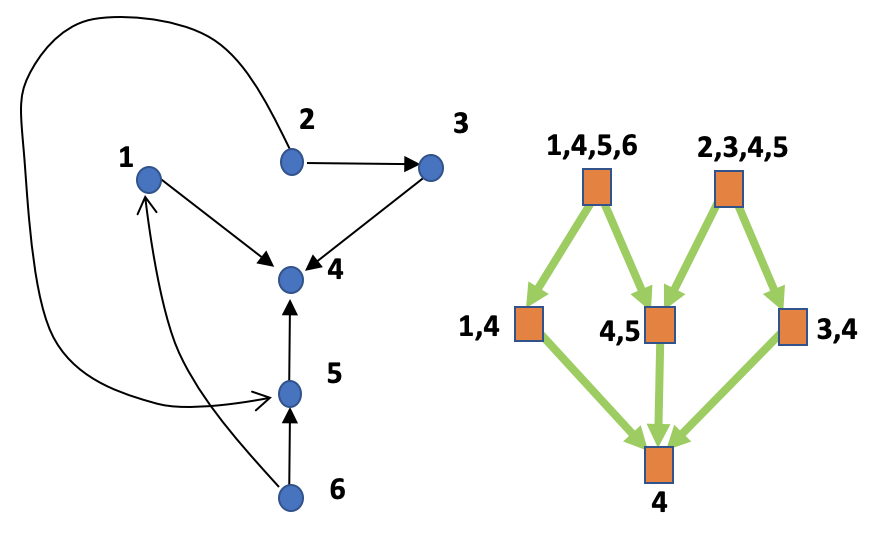}
\caption{a) The left panel indicates dependency constraints among a set of $6$ coordinates, using the same convention
as in Fig.\, 1. b) The right panel indicates an associated dependency graph, for units $A = \{4\}, B = \{1,4\}, C=\{4,5\}, D=\{3,4\},
E=\{1,4,5,6\}, F=\{2,3,4,5\}$.
}
\label{fig:2}
\end{figure}

To provide examples, consider the dependency graph illustrated in the right panel in \cref{fig:2}.
There are four associated type-1 distributions, $(1,0,0,0,0), (0,1,0,0,0,0),  (0,0,1,0,0,0), (0,0,0,1,0,0)$,
i.e., delta functions over the units $A, B, C,D$, respectively. There are six type-$2$ distributions; for example, the (unique) one
for unit $\oo = C$ is $\alpha = (1/2,0,1/2,0,0,0)$.

Next, note that units $A, B, C, E$ are all units contained in unit $E$ (i.e., those nodes are contained in the family
of node $E$). Similarly, $A, C$ are both units contained in $C$. Therefore
the units that are in $E$ but not in $C$ are $B, E$. Accordingly, 
the (unique) type-$3$ distribution for the pair of units $\oo = A, v = C$ is $\alpha =  (0,1/2,0,0,1/2,0)$.

$ $

\noindent II)
For any unit $\oo$, 
write $\MM(\oo) \subset \oo$ for any set of coordinates which can be represented by a unit structure 
$\MM(\oo)^* \subset \family(\oo)$~\footnote{There is always at least one 
such $\MM(\oo)$ for any non-leaf $\oo$, given by the set of coordinates $\cup \desc(\oo)$.
Note that in general, any such set of coordinates $\MM(\oo)$ can be represented by more than one unit $\MM(\oo) \subset \family(\oo)$.
For example, suppose there are leaves
 in ${\rm{desc}}(\oo)$ that have only one parent, i.e., there are units in ${\rm{desc}}(\oo)$ that are
only proper subsets of one other unit in ${\rm{desc}}(\oo)$. Then there is both a unit structure that contains those
leaves that represents $\MM(\oo)$, and a unit structure that does not contain those leaves but still represents $\MM(\oo)$.}.
Define $V^2(\NN^*)$ as the set of all distributions over $\B^{|\NN^*|}$ of the form
\eq{
\alpha = \dfrac{\sum_{\oo' \in \family(\oo) \setminus \MM(\oo)} \delta^{\oo'}}{\sum_{\oo' \in \family(\oo) \setminus \MM(\oo)} 1}
}
for some $\oo \in \NN^*$ and associated unit $\MM(\oo) \subset \family(\oo)$.  I will refer to any such distribution as a \textbf{type-$4$} distribution.
Note that any type-4 distribution $\alpha$ uniquely specifies both $\oo$ and $\MM(\oo)$. Given this, I will sometimes abuse notation
and write $\alpha^*$ for the unit structure with the single unit $\MM(\oo)$ specified by a type-4 distribution $\alpha$.

To illustrate type-$4$ distributions, again consider \cref{fig:2}. Choose $\oo = F$.
So $\family(\oo) = \{A, C, D, F\}$. Choose $\MM(\oo)$ to be the set of coordinates $\{3, 4, 5\}$ 
with the associated unit $\MM(\oo) = \{A,C, D\}$.
So $\family(\oo) \setminus \MM(\oo) = \{F\}$. The associated type-$4$ distribution is $\alpha = (0,0,0,0,0,1)$.


$ $

As shorthand, write $V(\NN^*) := V^1(\NN^*) \cup V^2(\NN^*)$. 
As a final piece of terminology, let $U = \{u\}$ be any set of distributions over some shared space. I will say that $U$ is \textbf{centered} if there exists a \textbf{centering distribution} $\pi \in \Delta_{U}$ 
such that $\E_\pi u = \sum_ {u \in U} u \pi_{u} $ equals the uniform distribution. 
Note that the set of all centering distributions of any $U$ is a convex polytope.

The following result involving centering distributions is proven in \cref{app:in_ex_EP_non-negative}:
\begin{proposition}
If $\pi$ is a centering distribution for $V(\NN^*)$, then
\eq{
\sigma^\NN \,\ge\, - \Delta \II^{\NN^*} -\, \sum_{\alpha \in V^2(\NN^*)}\pi_\alpha \Delta \II^{\alpha^*} \nonumber
}
\label{prop:1}
\end{proposition}

Note that the values of the in-ex informations $\II^{\NN^*}$ and $\Delta \II^{\alpha^*}$ 
at the beginning and end of the process are fully specified by $p_{t_i}$ and $p_{t_f}$. So given a centering distribution,
\cref{prop:1} provides a lower bound on global EP defined purely in terms of  $p_{t_i}$ and $p_{t_f}$. The precise rate matrix 
is irrelevant, so long as it has $\NN^*$ as a unit structure and maps $p_{t_i}$ to $p_{t_f}$.
Note as well that the sum in \cref{prop:1} only extends over distributions in $V^2(\NN^*)$. The role of the 
distributions in $V^1(\NN^*)$ is indirect; in general they are necessary to construct a centering distribution $\pi$,
and thereby constrain the possible values of $\pi_\alpha$ for the distributions  in $V^2(\NN^*)$.


As an example of \cref{prop:1}, first recall that even though
each unit $\oo$ evolves autonomously, in general we cannot choose all the rate matrices so that every $\sigma^\oo = 0$. In particular,
if $\oo$ has a non-trivial unit structure within it, then Eq.\,3.21
will apply for the choice $\NN = \oo$, potentially 
providing a strictly positive lower bound to $\sigma^\oo$. As a result, in general we cannot choose all the rate matrices so that
$ \SSS_{\oo \in {\NN^*}}\sigma^\oo = 0$, and so cannot in general lower bound $\sigma^\NN$ by $- \Delta \II^{\NN^*}$.

On the other hand, suppose that $\Gamma_{\NN^*}$ has height $2$. (So 
there are no units $\oo, \oo', \oo'' \in \NN^*$ such that $\oo'' \subset \oo' \subset \oo$.)
Then all delta function distributions over $\B^{|\NN^*|}$ are type-1 distributions, and so contained in $V^1(\NN^*)$.
Accordingly, the unit structure is centered by a distribution $\pi$ that is
uniform over all $\alpha \in {V}^1(\NN^*)$ and equals $0$ for all $\alpha \in V^2(\NN^*)$.
Plugging this into \cref{prop:1} establishes that the EP of any process with a height-2 unit structure $\NN^*$ is in fact lower-bounded 
by $- \Delta \II^{\NN^*}$.
This provides the formal justification for  Eq.\,4.1.

%
Moreover, in general, we can represent 
{any} process by using a unit structure of height $2$. (For example, we can
do that by combining all coordinates that are members of some unit $\oo$ that is not a root node of
$\Gamma_{\NN^*}$, into one, overarching unit.) Accordingly, for any
process, we can always find an associated unit structure $\NN^*$
for which $\sigma^\NN \ge - \Delta \II^{\NN^*}$. 

In addition, it is proven in \cref{app:always_centered} that for \textit{any} unit structure $\NN^*$, no matter what its height,
 $V(\NN^*)$ is centered. 
(The set of all associated centering distributions of $V(\NN^*)$ is
the convex polytope discussed in the introduction.)
In general, finding the optimal such centering distribution --- the one 
that maximizes the bound in \cref{prop:1}, and so provides the strongest lower bound on global EP --- only requires solving a linear programming problem. 

Unfortunately, there are some unit structures $\NN^*$ where the bound in \cref{prop:1} is negative
for an appropriate initial distribution $p_{t_i}(x)$ and conditional distribution $p(x({t_f}) \,|\, x(t_i))$ consistent with
$\NN^*$, no matter what centering distribution we use. In such cases, \cref{prop:1} does not provide a stronger bound on EP than the conventional second law.

On the other hand, as illustrated in the main text, often the bound in \cref{prop:1} will be stronger than the conventional second law.
Indeed, for \textit{every} unit structure $\NN^*$, and every associated centering distribution, 
there are initial distributions $p_{t_i}(x)$ and conditional distributions $p(x({t_f}) \,|\, x(t_i))$ that are consistent with
$\NN^*$ where the EP bound in \cref{prop:1} is at least as strong as the second law:
\begin{proposition}
Let $\NN^*$ be any unit structure that does not have $\NN$ itself as a member.
Then there exists an initial joint distribution $p_{t_i}(x)$ and a conditional distribution $p(x({t_f}) \,|\, x(t_i))$ consistent with $\NN^*$ such that for
any associated centering distribution $\pi_\alpha$,
\eq{
- \Delta \II^{\NN^*} - \sum_{\alpha \in V^2(\NN^*)}\pi_\alpha \Delta \II^{\alpha^*} \ge 0
}
for every rate matrix that both implements that $p(x({t_f}) \,|\, x(t_i))$ and obeys the unit structure. 
\label{prop:2}
\end{proposition}
\noindent (See \cref{app:negative_in_ex_info} for proof.)

In addition, even if $- \Delta \II^{\NN^*} < 0$, often we can use Eq.\,3.21
directly to provide a non-negative lower bound
on the global EP, by constructing a lower bound on $\SSS_{\oo \in {\NN^*}}\sigma^\oo$ which is bigger than $- \Delta \II^{\NN^*}$.
To construct such a bound, first note that since there is no unit structure within leaf units of the dependency graph, in theory
we can implement all such units with zero local EP. Next, for any unit $\oo$ that is a parent of a leaf unit, we can often use
Eq.\,15 in~\cite{wolpert2020minimal} (or its corollary, Eq.\,17) to construct strictly positive lower bounds on $\sigma^\oo$.

Write the polytope of all centering distributions of the unit structure $\NN^*$ as ${\mathcal{P}}(\NN^*)$, and write the strongest lower bound on EP given by \cref{prop:1} as
\eq{
&{\mathcal{F}}\left(p_{t_i}, \, p_{t_f}, \, \NN^*\right) \nonumber \\
&\qquad := -\min_{\pi \in {\mathcal{P}}(\NN^*)}  \left[\Delta \II^{\NN^*} + \sum_{\alpha \in V^2(\NN^*)}\pi_\alpha \Delta \II^{\alpha^*}\right]
}
Combining \cref{prop:1,prop:2} establishes that for any unit structure $\NN^*$
there are pairs of $p_{t_i}(x)$ and a conditional distribution $p(x({t_f}) \,|\, x(t_i))$ consistent with $\NN^*$ such that 
\eq{
Q + \Delta S & \ge\;  {\mathcal{F}}\left(p_{t_i}, p_{t_f}, \, \NN^*\right)  \\
 	&\ge 0
\label{eq:19}
}
In contrast, the conventional second law says only that $Q + \Delta S \ge 0$, no matter what $p_{t_i}$, $p_{t_f}$ or $\NN^*$ are.

Summarizing, suppose we are given a unit structure $\NN^*$ that applies to the rate matrix at all times, an initial distribution $p_{t_i}(x)$, and 
a conditional distribution $p\left(x({t_f}) \,|\, x(t_i)\right)$ consistent with $\NN^*$.  
Then we know that $- \Delta \II^{\NN^*} - \sum_{\alpha \in V^2(\NN^*)}\pi_\alpha \Delta \II^{\alpha^*}$ is a lower bound on the EP,
for any $\pi$ that is a centering distribution for $\NN^*$. 
These lower bounds are simple to evaluate, and are often stronger than the second law.
We can find the strongest such lower bound on EP due to the unit structure $\NN^*$ by solving a linear programming problem. 

Furthermore, in general, it is possible to represent any given set of constraints with more than one
unit structure. Each one of them results in its own strongest lower bound on EP, given by solving the associated linear programming problem.   
So to find the strongest version of the second law for a given set of constraints, we should solve all the linear programming problems
specified by the unit structures that can represent those constraints.

This result extends the previous strengthenings of the second law
derived in~\cite{wolpert_thermo_comp_review_2019,kolchinsky_wolpert_2020_thermo_under_constraints,Boyd:2018aa}, which
all assume that the units have no overlap, to the case
where the units may overlap with one another in arbitrary ways, even if none of the coordinates are fixed in the dynamics.
In addition to this result, which maps an arbitrary set of constraints,  a $p_{t_i}$ and a $p_{t_f}$ to a lower bound on EP, a second
result is that for any set of constraints, there is a $p_{t_i}$ and $p_{t_f}$ that results in a strictly positive lower bound on EP,
stronger than the conventional second law.


\section{Proof of \cref{prop:1}}
\label{app:in_ex_EP_non-negative}

The proof has two parts.
First, I construct a function $f : \NN \rightarrow \R$ such that for all units $\oo \in \NN^*$, 
$\sum_{i \in \oo} f_i = \sigma^\oo$. (Note that in general, any coordinate $i$ will
be in more than one unit $\oo$, and so this function $f$ is the solution to a set of coupled equations.) I will
then apply the inclusion-exclusion principle with this $f$ in order to replace the first in-ex sums over unit $\oo$ on the RHS of 
Eq.\,3.21
with a conventional sum over coordinates, $\sum_i f_i$. 

In the second part of the proof I use the hypothesized existence of a centering
distribution to provide a lower bound on $\sum_i f_i$, expressed purely in
terms of $p_{t_i}$ and $p_{t_f}$. Plugging in to Eq.\,3.21
then completes the proof.

To begin, for each unit $\oo$, define $\overline{\oo}$ as the set of all coordinates in $\oo$
that are not in any of the units in $\desc(\oo)$. (As an example, in Fig.\,1 in the main text,
$\overline{\oo}$ is the pair of coordinates $1$ and $2$.) Because ${\NN^*}$ is closed
under intersections and covers $\NN$, every coordinate is in $\overline{\oo}$ for exactly one unit $\oo \in \NN^*$.
Moreover, 
because there are no vacuous units allowed,
$\overline{\oo}$ is nonempty for every unit $\oo$.
Note that if $\oo \subset \oo'$ for two units $\oo, \oo'$, then the coordinates
in $\oo$ must evolve independently of the states of any coordinates in $\overline{\oo'}$, but the
reverse need not be true. In other words, if there is an edge from $\oo'$ to $\oo$, then 
there may be coordinates in $\overline{\oo'}$  whose dynamics depends on the state of coordinates in $\oo$, but not vice-versa.

For all $j \in \N$, let $\Omega_j$ be the set of all nodes in $\Gamma_{\NN^*}$, with height $j$. So in particular, $\Omega_1$ is the the set all units with no subunits.
For every $\oo \in \Omega_1$, for all coordinates $i \in \oo$, set 
\eq{
f_i := \dfrac{\sigma^\oo} {|\oo|}
}
So by construction, for all $\oo \in \Omega_1$, 
\eq{
\sum_{i \in \oo} f_i &= \sigma^{\oo}
\label{eq:a1}
}
Since  $\sigma^\oo \ge 0$ for all $\oo \in \NN^*$, this means that $\sum_{i \in \oo} f_i \ge 0$ for all $\oo \in \Omega_1$.


Next, note that any for any $\oo \in \NN^*$,
the set of units  in  $\desc(\oo)$ is closed under intersections.
This allows us to 
define
\eq{
f_i := \dfrac{1}{|\overline{\oo}|} \left(\sigma^\oo - \widehat{ \sum_{\oo' \in \desc(\oo)}} \sigma^{\oo'}\right)
\label{eq:a2}
}
for all $j \in \NN$, all $\oo \in \Omega_j$, and all $i \in \overline{\oo}$.
(Note that any such $i$ will be assigned a value $f_i$ exactly once in this procedure.)

In general, it could be that
$f_i$ is negative. Note though that since no two units in $\Omega_1$ have any overlap, for all $\oo \in \Omega_2$,
\eq{
\widehat{ \sum_{\oo' \in \desc(\oo)}} \sigma^{\oo'} &=  \widehat{ \sum_{\oo' \in \desc(\oo)}} \sum_{i \in \oo'} f_i \\
	&= \sum_{i \in \cup \desc(\oo)} f_i
}
Therefore  by \cref{eq:a2}, for all $\oo \in \Omega_2$,
\eq{
\sum_{i \in \oo} f_i &= \sum_{i \in \cup \desc(\oo)} f_i + \sum_{i \in \overline{\oo}} f_i \\
	&=  {\widehat{ \sum_{\oo' \in \desc(\oo)}}} \sigma^{\oo'}  + \sum_{i \in \overline{\oo}} f_i \\
	&= \sigma^\oo
\label{eq:a4}
}
i.e., \cref{eq:a1} holds for unit $\oo$.

Next, assume that \cref{eq:a1} holds for all $\oo \in \Omega_{k-1}$ for some integer $k > 2$. 
Then  by \cref{eq:a2} and the inclusion-exclusion principle, for any $v \in \Omega_k$,
\eq{
\sum_{i \in v} f_i &= \sum_{i \in \cup \desc(v)} f_i + \sum_{i \in \overline{v}} f_i \\
	&=  {\widehat{ \sum_{\oo' \in \desc(v)}}} \sum_{i \in \oo'} f_i  + \sum_{i \in \overline{v}} f_i \\
	&= {\widehat{ \sum_{\oo' \in \desc(v)}} }\sigma^{\oo'} + \sum_{i \in \overline{v}} f_i \\
	&= \sigma^v
\label{eq:a12}
}
(Note that the inclusion-exclusion principle holds for arbitrary functions $f$, not
just for nowhere-negative functions.) 
Since $\sigma^v \ge 0$, this means that we are guaranteed that $\sum_{i \in v} f_i \ge 0$.

Iterate this procedure going from nodes in $\Omega_{k-1}$ to those in $\Omega_k$ 
until all units have been considered, so that values $f_i$ have been assigned to all coordinates $i \in \NN$.
By induction, at the end of this procedure,  for all units $\oo \in \NN^*$, \cref{eq:a1} will hold and
$\sum_{i \in \oo} f_i \ge 0$. In addition, by the inclusion-exclusion principle,
\eq{
\SSS_{\oo \in \NN^*} \sigma^\oo &= \SSS_{\oo \in \NN^*} \left[\sum_{i \in \oo} f_i\right] \\
	&=  \sum_{i \in \NN} f_i 
\label{eq:a14}
}
(Note that since $\NN\not \in \NN^*$, \cref{eq:a12} does not imply that $\sum_{i \in \NN} f_i = \sigma^\NN$. So we cannot
combine \cref{eq:a14} with the fact that global EP is $\ge 0$
to establish that $\SSS_{\oo \in \NN^*} \sigma^\oo$ is also non-negative.)

It will be convenient to define new variables that equal sums of $f_i$ over small sets of coordinates $i$. For all $\oo \in \NN^*$, define
\eq{
\label{eq:g_def}
g_\oo &:= \sum_{i \in \overline{\oo}} f_i \\
	&= \sigma^\oo - \widehat{ \sum_{\oo' \in \desc(\oo)}} \sigma^{\oo'}
\label{eq:g_def2}
}
where the second line follows from \cref{eq:a2}.
Since each coordinate $i$ is in $\overline{\oo}$ for exactly one unit $\oo$,
for any $v \in \NN^*$,
\eq{
g_v \;+\!\! \sum_{\oo \in \desc{(v)}} g_\oo &= \sum_{i \in \overline{v}} f_i  \;+\!\!  \sum_{\oo \in \desc{(v)}} \sum_{i \in \overline{\oo}} f_i  \\
\label{eq:g_sum}
	&=  \sum_{i \in v} f_i  \\
	&= \sigma^v
\label{eq:a19a}
}
where the last line uses \cref{eq:a12}.
Using similar reasoning shows that $\sum_{\oo \in \NN^*} g_\oo = \sum_{i \in \NN} f_i$. Combining this with \cref{eq:a14}
and \cref{eq:12} gives
\eq{
\sigma^\NN + \Delta \II^{\NN^*} = \sum_{\oo \in \NN^*} g_\oo
\label{eq:a20}
}

This completes the first part of the proof. In the second part I derive a lower bound on
the RHS of \cref{eq:a20}. First, to reduce the complexity of the equations, I will translate all distributions $\alpha$ into binary-valued
vectors:
\begin{enumerate}
\item ${\hat{\alpha}} = \alpha$ for any type-1 distribution $\alpha$;
\item ${\hat{\alpha}} = \alpha{\sum_{\oo' \in \family(\oo)} 1}$ for any type-2 distribution $\alpha$;
\item ${\hat{\alpha}} = \alpha{\sum_{\oo' \in \family(\oo) \setminus \family(v)} 1}$ for any type-3 distribution $\alpha$;
\item ${\hat{\alpha}} = \alpha{\sum_{\oo' \in \family(\oo) \setminus \MM^*(\oo)} 1}$ for any type-4 distribution $\alpha$;
\end{enumerate}
Note that every component of every vector $\hat{\alpha}$ is either a $0$ or a $1$.
I will refer to any vectors that obey condition (1) as \textbf{type-$1$} vectors, and similarly for vectors obeying conditions (2), (3) and / or (4).

Now make three suppositions. First, suppose that for all vectors ${\hat{\alpha}}$ of types $1, 2$ or $3$,
\eq{
\sum_{\oo  \in \NN^*} g_\oo {\hat{\alpha}}_\oo \ge 0
\label{eq:a21}
}
Next, suppose that for all vectors ${\hat{\alpha}}$ of type-$4$, and all associated unit structures ${\hat{\alpha}}^*$,
\eq{
\sum_{\oo  \in \NN^*} g_\oo {\hat{\alpha}}_\oo \ge -\Delta \II^{{\hat{\alpha}}^*}
\label{eq:a22}
}
Now, hypothesize  that there is a \textbf{centering vector} $\gamma$ all of whose components are non-negative such that
\eq{
\sum_{{\hat{\alpha}}  \in V^1(\NN^*)} \gamma_{\hat{\alpha}} {\hat{\alpha}}_\oo + \sum_{{\hat{\alpha}}  \in V^2(\NN^*)}\gamma_{\hat{\alpha}} {\hat{\alpha}}_\oo = 1
\label{eq:a23}
}
for all $\oo \in \NN^*$.
Multiplying both sides of  \cref{eq:a23} by $ g_\oo$, summing over $\oo$, and then plugging in \cref{eq:a21,eq:a22},
we see that if those three equations hold,
\eq{
 \sum_{\oo \in \NN^*} g_\oo \;\ge\; -\sum_{{\hat{\alpha}}  \in V^2(\NN^*)}\gamma_{\hat{\alpha}} \Delta \II^{{\hat{\alpha}}^*}
}
Plugging this into \cref{eq:a20} shows that if we can prove that the suppositions \cref{eq:a21,eq:a22} always hold, then we will have proven that
for any centering vector $\gamma$,
\eq{
\sigma^\NN &\ge - \Delta \II^{\NN^*} - \sum_{{\hat{\alpha}} \in V^2(\NN^*)}\gamma_{\hat{\alpha}} \Delta \II^{{\hat{\alpha}}^*}
}

To begin, use \cref{eq:g_def2} to conclude that $g_\oo = \sigma^\oo$ for all $\oo \in \Omega_1$ (which have no descendants) and so
$g_\oo \ge 0$ for all  $\oo \in \Omega_1$. Next, combine this fact that $g_\oo = \sigma^\oo$ for all leaf nodes $\oo$ with 
\cref{eq:13a,eq:a19a} to also conclude that $g_\oo \ge 0$ for all $\oo \in \Omega_2$.
Combining these two results means that \cref{eq:a21} holds for all type-1 vectors ${\hat{\alpha}}$.

Next, note that \cref{eq:a19a} means that for any $\oo \in \NN^*$,
\eq{
\sum_{\oo' \subseteq \oo} g_{\oo'} = \sigma^\oo
}
So by the non-negativity of local EP, for all $\oo \in \NN^*$, 
\eq{
\sum_\oo \sum_{\oo' \subseteq \oo} \delta(\oo', \oo) g_{\oo} \ge 0
}
This means that \cref{eq:a21} holds for all type-2 vectors ${\hat{\alpha}}$.

Now consider any pair of nodes $\oo \in \NN^*, v \subset \oo$. Using \cref{eq:a19a}
for both $\oo$ and $v$ and then applying Eq.\,3.15
establishes that
\eq{
0 &\le \sum_{\oo' \in \family(\oo)} g_{\oo'}- \sum_{\oo' \in \family(v)} g_{\oo'}  \\
	&= \sum_{\oo' \in \family(\oo) \setminus \family(v)} g_{\oo'}
}
This means that \cref{eq:a21} also holds for all type-3 vectors ${\hat{\alpha}}$.
Combining establishes our first goal, of showing that \cref{eq:a21} holds for all vectors ${\hat{\alpha}} \in V^1(\NN^*)$,
of types $1, 2$ or $3$.

Next, consider any pair of a unit $\oo$ and set of coordinates $\MM(\oo) \subset \oo$ such that there is a unit structure $\MM(\oo)^* \subset \family(\oo)$. 
Use \cref{eq:a12}, the inclusion-exclusion principle, \cref{eq:g_def}, and then \cref{eq:a19a} to expand
\eq{
\sigma^\oo - \SSS_{\oo' \in {\MM(\oo)^*}}\sigma^{\oo'} &= \sigma^\oo - \SSS_{\oo' \in {\MM(\oo)^*}} \left[\sum_{i \in \oo'} f_i\right] \\
	&=  \sigma^\oo - \sum_{i \in \MM(\oo)} f_i \\
	&=  \sigma^\oo - \sum_{\oo'\in \MM(\oo)^*} \sum_{i \in \overline{\oo'}} f_i \\
	&=  \sigma^\oo - \sum_{\oo' \in \MM(\oo)^*} g_{\oo'}\\
	&=  g_\oo + \sum_{\oo' \in \desc(\oo)} g_{\oo'} - \sum_{\oo' \in \MM(\oo)^*} g_{\oo'} \\
	&=  g_\oo + \sum_{\oo' \in \desc(\oo) \setminus \MM(\oo)^*} g_{\oo'} \\
	&=  \sum_{\oo' \in \family(\oo) \setminus \MM(\oo)^*} g_{\oo'} 
}
Eq.\,3.22 then establishes that 
\eq{
 \sum_{\oo' \in \family(\oo) \setminus \MM(\oo)^*} g_{\oo'}  \ge  -\Delta \II^{\MM(\oo)^*}
}
Plugging this into the definition of type-$4$ vectors for ${\hat{\alpha}} = \MM(\oo)$ and
${\hat{\alpha}} = \MM(\oo)^*$ confirms that \cref{eq:a21} holds. 


Combining establishes that for any centering vector $\gamma$, 
\eq{
\sigma^\NN &\ge - \Delta \II^{\NN^*} - \sum_{{\hat{\alpha}} \in V^2(\NN^*)}\gamma_{\hat{\alpha}} \Delta \II^{{\hat{\alpha}}^*}
\label{eq:a38}
}
Finally, normalize each vector $\hat{\alpha}$ to recover the distributions $\alpha$,
and define the distribution $\pi(\alpha)$ by normalizing $\gamma(\alpha)$. It follows that $\E_\pi \alpha$ is
the uniform distribution,
so that $\pi$ is a centering distribution. In addition,
\cref{eq:a38} gets converted into the bound in \cref{prop:1}. This completes the proof of \cref{prop:1}.

\section{Proof that any unit structure is centered}
\label{app:always_centered}

To begin, choose $V^1(\NN^*)$ to be the set of all type-1 distributions, i.e., $V^1(\NN^*)$ is the set of all distributions $\delta^\oo$ for any $\oo$ of height $\le 2$. 

Next, for each $\oo$ of height greater than $2$, plug $\MM(\oo) = \desc(\oo)$ and any arbitrary single
one of the possible unit structures $\MM(\oo)^*$ into \cref{eq:a23} to define a distribution 
\eq{
\alpha(\oo) &= \sum_{\oo' \in \family(\oo) \setminus \MM^*(\oo)} \delta^{\oo'}  \\
	&= \delta^\oo
}
and associated unit structure $\alpha(\oo)^*$.
Choose $V^2(\NN^*)$ to be the set of all such $\alpha(\oo)$, one per $\oo$, as one ranges over all $\oo$ of
height greater than $2$.

By construction, $V(\NN^*)$ is exactly the set of all vectors $\delta^\oo$ as one ranges over all $\oo \in \NN^*$.
Accordingly, the sum of all distributions in $V(\NN^*)$ is $\vec{1}$, and the average of those
distributions is the uniform distribution, $(1 / |\NN^*|, 1 / |\NN^*|, \ldots)$. Therefore the set of those
vectors is centered, where the centering distribution $\pi_\alpha = 1 / |\NN^*|$ 
for all $\alpha \in V(\NN^*)$. This completes the proof.

\section{Proof of \cref{prop:2}}

\label{app:negative_in_ex_info}

First, note that for a uniform distribution over the states of the multi-dimensional system, the entropy of every coordinate $i$ with $|X_i|$ states is $\ln |X_i|$. Furthermore, no matter what the unit
structure is, one can create a process consistent with that structure that results in this uniform
distribution as the final distribution. (Just choose the rate matrix so that by the end of the process,
for each coordinate $i$, $x_i$  has been uniformly randomized.) 

Now assign values $f_i =  \ln |X_i|$ to all coordinates. By construction, for all units $\oo$, $\sum_{i \in \oo} f_i =  S^\oo$.
In addition, 
\eq{
\SSS_{\oo} \left[\sum_{i \in \oo} f_i\right] = \sum_i f_i 
}
by the inclusion-exclusion principle. 
But the sum on the RHS just equals $S^\NN$, the entropy of the full system, since the coordinates are statistically independent under
the final distribution. Therefore $\II^{\NN^*} = 0$ for this ending distribution. Similarly, for this ending distribution,
$\II^{\alpha^*} = 0$ for every $\alpha \in V^2(\NN^*)$.

So to find a situation where \cref{prop:1} holds, it suffices 
to find an initial distribution $p$ such that $\II^{\NN^*}(p) \ge 0$ while $\II^{\alpha^*}(p) = 0$ for all $\alpha \in  V^2(\NN^*)$. To do 
that, label the states of each coordinate $i$ by the first $|X_i|$ counting numbers. 
Define $M := \min_{i\in \NN} |X_i|$, and define $\Gamma_{\NN^*}^R$ as the (units corresponding to the) root nodes of the dependency graph $\Gamma_{\NN^*}$.
Note that since by hypothesis $\NN \not\in \NN^*$, there must be at least two distinct root nodes in $\Gamma_{\NN^*}^R$.
Furthermore, since there are no vacuous units in a composite
system, all of (the units corresponding to) those root nodes contain coordinates that are not in
any other units, i.e., for all $\oo \in \Gamma_{\NN^*}^R$, $\overline{\oo} \ne \varnothing$.

Next, define $T :=  \cup_{\oo \in \Gamma_{\NN^*}^R} \overline{\oo}$, and fix the state of each coordinate $j \not \in T$,
i.e., set the distribution over the state of that coordinate to a delta function. Set the joint distribution
over the remaining coordinates to
\eq{
p(x_{T}) &=  \dfrac{1}{M}\sum_{k = 1}^{M}\prod_{i\in T} \delta(x_i, k)
\label{eq:b2}
}
So all coordinates that only occur in a single root unit are perfectly coupled with one another, with a uniform distribution over the set of $M$ possible
joint states they can adopt. 

The entropy of the full joint distribution defined this way is $S(p) = \ln M$. So to prove that $\II^{\NN^*}(p) \ge 0$ we need to show that 
\eq{
\SSS_{\oo} S(p_\oo) &\ge \ln M
}
To do that, assign the value $f_j = 0$ to all coordinates $j \not \in T$. For each
coordinate $j \in T$, where $\oo(j)$ is the unique unit containing $j$, assign the value
\eq{
f_j &= \dfrac{\ln M} {|\overline{\oo(j)}| }
}
where $|\overline{\oo(j)}|$ is the number of elements in $\overline{\oo(j)}$.

By construction, for all units $\oo \in \NN^*$,
\eq{
S(p_\oo) = \sum_{j \in \oo} f_j
}
Accordingly, by the inclusion-exclusion principle
\eq{
\SSS_{\oo} S(p_\oo) &= \sum_{i \in \NN} f_i \\
	\nonumber \\
	&= \sum_{\oo \in \Gamma_{\NN^*}^R} \ln M
}
Since $\Gamma_{\NN^*}^R$ contains at least two units, this means that
\eq{
\SSS_{\oo}S(p_\oo) & > \ln M = S(p)
}
In addition, the entropy of every unit $\oo \not \in \Gamma_{\NN^*}^R$ equals $0$. Accordingly,
$\II^{\alpha^*}(p) = 0$ for all $\alpha \in  V^2(\NN^*)$. 

This completes the proof.

\section{Proofs of results for thermodynamics of feedback control of composite unit structures}

\label{app:proofs_for_feedback_scenario}

The growth of the entropy of unit $\oo$ during the process changes when we 
expand it into the unit $\oo'(\oo)$ that includes $C$. To calculate how much it changes,
first, since $\CC$ does not change state during the process, the entropy of each unit $\oo'(\oo)$ grows during the process by
the change in conditional entropy,
\eq{
S(p_{t_f}(X_{\oo} | C)) - S(p_{t_i}(X_{\oo} | C))
\label{eq:j.1}
}
where $p_{t_i}, p_{t_f}$ are given by Eq.\,5.1, 5.2,
respectively.
In contrast, the growth of entropy in the unit $\oo$ in the original, no-feedback process is
\eq{
S(p^\dagger_{t_f}(X_{\oo})) - S(p^\dagger_{t_i}(X_{\oo}))
}
This just equals $S(p_{t_f}(X_{\oo})) - S(p_{t_i}(X_{\oo} ))$, due to our 
assumption that the initial and final marginal distributions over $X$  are the same regardless of whether
there is a feedback apparatus.

Combining, we see that the 
change in the growth of entropy of unit $\oo$ when we add the feedback apparatus is the drop of mutual information,
\eq{
I_{p_{t_i}}(X_{\oo} ; C) - I_{p_{t_f}}(X_{\oo} ; C) = -\Delta I(X_\oo ; C)
}
This is true for every unit $\oo \in \NN^*$, and for $\NN$ itself. Plugging this fact into Eq.\,4.1 gives Eq.\,5.4.

Next, reuse the reasoning behind \cref{eq:j.1} to establish that the lower bound on EP in the feedback scenario is
\eq{
 \Delta S(X_\NN | C) - \Delta \left[\SSS_{\oo \in {\NN^*}}  S(X_\oo | C) \right] 
\label{eq:1.9}
}
In addition, the maximal work that can be extracted from the system
under feedback control \textit{without consideration of the unit structure of the system} is~\cite{sagawa2009minimal,parrondo2015thermodynamics, kolchinsky_wolpert_2020_thermo_under_constraints}
\eq{
-\Delta F(X_\NN) - \Delta I(X_\NN; C) = \Delta S(X_\NN) - \Delta I(X_\NN; C) = \Delta S(X_\NN | C)
\label{eq:1.10}
}
(under the common assumption that the Hamiltonian at both $t_i$ and $t_f$ is uniform, and in units
of $k_B T = 1$).
Subtracting the lower EP bound, \cref{eq:1.9}, from \cref{eq:1.10}  establishes Eq.\,5.5.

\section{Proof of \cref{prop:1a}}
\label{app:prop_1}

By \cref{eq:2.5},
\eq{
K^{x'_\oo}_{x_\oo}(t) 
	&= \sum_{x_{-\oo}} K^{x'_\oo, x'_{- \oo}}_{x_\oo, x_{-\oo}}(t) \\
	&= \sum_{x_{-\oo}} \sum_{v }  L^{x'_{\LL(v)}}_{x_{\PP(v)}}(v; t)  \delta^{x'_{_{-\PP(v)}}}_{x_{_{-\PP(v)}}} \\
	&=   \sum_{v } \sum_{x_{-\oo}}
		L^{x'_{\LL(v) \cap \oo},  x'_{\LL(v) \setminus \oo}}_{x_{\PP(v) \cap \oo}, x_{\PP(v) \setminus \oo}}(v;t)
		\delta^{x'_{_{-\PP(v)}}}_{x_{_{-\PP(v)}}}  \\
	&=   \sum_{v \in \nu(\oo)} \sum_{x_{-\oo}}
		L^{x'_{\LL(v) \cap \oo},  x'_{\LL(v) \setminus \oo}}_{x_{\PP(v) \cap \oo}, x_{\PP(v) \setminus \oo}}(v;t)
		\delta^{x'_{_{-\PP(v)}}}_{x_{_{-\PP(v)}}}   \\
	&=   \sum_{v \in \nu(\oo)} \sum_{x_{-\oo}}
		L^{x'_{\LL(v) \cap \oo},  x'_{\LL(v) \setminus \oo}}_{x_{\PP(v) \cap \oo}, x_{\PP(v) \setminus \oo}}(v;t)
		\delta^{x'_{_{\oo \setminus \PP(v)}}}_{x_{_{\oo \setminus \PP(v)}}} 
\label{eq:2.19}
}
where 
the penultimate line sums the Kronecker delta function over all $x_{-\PP(v) \cap -\oo}$, and the last line uses
the fact that for all $v$ such that $\PP(v) \cap \oo = \varnothing$, the inner sum over $x_{-\oo}$ equals zero, by normalization. 

Using the facts that \cref{eq:2.19} must be independent of $x'_{-\oo}$, $\PP(v) \setminus \oo \subseteq \LL(v) \setminus \oo$,
and a sum of rate matrices is a rate matrix, we can rewrite \cref{eq:2.19} as
\eq{
K^{x'_\oo}_{x_\oo}(t) 	& = \sum_{v \in \nu(\oo)} \sum_{x_{_{-\oo\cap -\PP(v)}}}
		\underline{L}^{x'_{\LL(v) \cap \oo}}_{x_{\PP(v) \cap \oo}}(v, \oo;t) 
		\delta^{x'_{_{\oo \setminus \PP(v)}}}_{x_{_{\oo \setminus \PP(v)}}}   \\
	&=  \sum_{v \in \nu(\oo)} \sum_{x_{_{-(\oo \cup \PP(v))}}}
		\underline{L}^{x'_{\LL(v) \cap \oo}}_{x_{\PP(v) \cap \oo}}(v, \oo;t) 
		\delta^{x'_{_{\oo \setminus \PP(v)}}}_{x_{_{\oo \setminus \PP(v)}}}  \\
	&= \big(N- |\oo \cup \PP(v)|\big) \sum_{v \in \nu(\oo)} \underline{L}^{x'_{\LL(v) \cap \oo}}_{x_{\PP(v) \cap \oo}}(v, \oo;t)
		\delta^{x'_{_{\oo \setminus \PP(v)}}}_{x_{_{\oo \setminus \PP(v)}}}
}
where $x'_{\LL(v) \setminus \oo}$ is arbitrary and
\eq{\underline{L}^{x'_{\LL(v) \cap \oo}}_{x_{\PP(v) \cap \oo}}(v, \oo;t) &:=
	\sum_{x_{_{\PP(v) \setminus \oo}}} L^{x'_{\LL(v) \cap \oo},  x'_{\LL(v) \setminus \oo}}_{x_{\PP(v) \cap \oo}, x_{\PP(v) \setminus \oo}}(v;t) 
}
Setting 
\eq{
	\widehat{L}^{x'_{\LL(v) \cap \oo}}_{x_{\PP(v) \cap \oo}}(v, \oo;t)  &=	
			\big(N- |\oo \cup \PP(v)|\big) \underline{L}^{x'_{\LL(v) \cap \oo}}_{x_{\PP(v) \cap \oo}}(v, \oo;t) 
}
completes the proof.

\section{Proof of \cref{prop:1b}}
\label{app:proof_prop:1b}

In sequence, use \cref{eq:2.10,eq:2.5}, the facts that $\PP(\oo) = \oo, \PP(-\oo) = -\oo$ and 
that the matrices $L(v;t)$ are all normalized, and then \cref{eq:2.8} to expand
\eq{
K^{x'_\oo}_{x_\oo}(\oo; t) &= \sum_{x_{-\oo}} K^{x'_\oo, x'_{-\oo}}_{x_\oo, x_{-\oo}}(t) \\
	&=  \sum_{x_{-\oo}} \sum_v L^{x'_\oo, x'_{-\oo}}_{x_\oo, x_{-\oo}}(v; t) \\
	&=  \sum_{x_{-\oo}}\left( \sum_{v \in \nu(\oo)} L^{x'_\oo, x'_{-\oo}}_{x_\oo, x_{-\oo}}(v; t) 
			+  \sum_{v \not\in \nu(\oo)} L^{x'_\oo, x'_{-\oo}}_{x_\oo, x_{-\oo}}(v; t)\right) \\
	&=  \sum_{x_{-\oo}}\left( \sum_{v \in \nu(\oo)} L^{x'_\oo, x'_{-\oo}}_{x_\oo, x_{-\oo}}(v; t) 
			+  \sum_{v \in \nu(-\oo)} L^{x'_\oo, x'_{-\oo}}_{x_\oo, x_{-\oo}}(v; t)\right) \\
	&=  \sum_{x_{-\oo}} \sum_{v \in \nu(\oo)} L^{x'_\oo, x'_{-\oo}}_{x_\oo, x_{-\oo}}(v; t)  \\
	&=  \sum_{x_{-\oo}} \sum_{v \in \nu(\oo)} L^{x'_{\LL(v)}}_{x_{\PP(v)}}(v; t) \delta^{x'_{-\PP(v)}}_{x_{-\PP(v)}} \\
	&= 	\sum_{v \in \nu(\oo)} L^{x'_{\oo}}_{x_{\oo}}(v; t)
}
which completes the proof.

\bibliographystyle{amsplain}
\bibliography{../../../../BIB/thermo_refs.main.3.BIB.DIR,new.motivation.strengthened.second.law.rsta.version.main.POST.RESPONSE.TO.REFEREES.v2.bib}

\end{document}